\documentclass[12pt,preprint]{aastex}

\shorttitle{Scaling Mass Profiles around Elliptical Galaxies}
\shortauthors{Fukazawa et al.}

\received{2005 April 29}
\begin{document}


\title{Scaling Mass Profiles around Elliptical Galaxies Observed with
{\it Chandra} and {\it XMM-Newton}}


\author{Y. Fukazawa\altaffilmark{1}, J. G. Betoya-Nonesa\altaffilmark{1}, J. Pu\altaffilmark{1}, A. Ohto\altaffilmark{1}, and N. Kawano\altaffilmark{1}}
\affil{Department of Physical Science, School of Science,
Hiroshima University, 1-3-1 Kagamiyama, Higashi-Hiroshima, Hiroshima 739-8526}

\email{fukazawa@hirax6.hepl.hiroshima-u.ac.jp}






\begin{abstract}
We investigated the dynamical structure of 53 elliptical galaxies, based
on the {\it Chandra} archival X-ray data.
In X-ray luminous galaxies, a temperature increases with radius 
and a gas density is systematically higher at the optical outskirts, 
indicating a presence of a significant amount of the group-scale hot gas.
In contrast, X-ray dim galaxies show a flat or declining 
temperature profile against radius and the gas density is relatively 
lower at the optical outskirts.
Thus it is found that X-ray bright and faint elliptical galaxies are clearly
 distinguished by the temperature and gas density profile.
The mass profile is well scaled by a virial radius $r_{200}$ rather than
 an optical-half radius $r_e$, and is quite similar at
 $(0.001-0.03)r_{200}$ between X-ray luminous and dim galaxies,
and smoothly connects to those of clusters of galaxies.
At the inner region of $(0.001-0.01)r_{200}$ or $(0.1-1)r_e$, 
the mass profile well traces a stellar mass with a constant
mass-to-light ratio of $M/L_{\rm B}=3-10(M_{\odot}/L_{\odot})$. 
$M/L_{\rm B}$ ratio of X-ray bright galaxies rises up steeply beyond 
$0.01r_{200}$, 
and thus requires a presence of massive dark matter halo.
From the deprojection analysis combined with the {\it XMM-Newton} data, 
we found that 
X-ray dim galaxies, NGC 3923, NGC 720, and IC 1459, also 
have a high $M/L_{\rm B}$ ratio of
20--30 at 20 kpc, comparable to that of X-ray luminous galaxies.
Therefore, dark matter is indicated to be common in elliptical galaxies,
 and their distribution almost follows the NFW profile, as well as
 galaxy clusters.
\end{abstract}




\section{Introduction}

Dynamical studies indicate the presence of dark matter (DM) around spiral
galaxies \citep{persic96}, and 
the DM mass profile has been measured through the rotation
velocity profile.
On the other hand, random motions of stars and clouds did not allow us to
determine accurately the mass profile in elliptical galaxies.
Alternatively, X-ray distribution of a hot interstellar medium (ISM) traces 
the gravitational potential \citep{forman85}, 
and, using ROSAT and ASCA satellites, it is revealed that 
massive DM halo exists around X-ray bright giant elliptical galaxies
at (5--10)$r_e$, where $r_e$ is an effective radius or optical half-light
radius \citep{ikebe96, trinchieri97, loewenstein99, matsushita98, 
matsushita01}.
Numerical simulations based on cold cark matter (CDM) predict the cusp
structure of the density profile \citep{navarro97}, whose shape is well
scaled for various systems from small galaxies to rich clusters.
The DM halo structure is sensitive to the properties of the DM, and
attention of many workers is now focussed on the inner slope of the DM
density profile.
DM profile of galaxy clusters measured with {\it Chandra} and
{\it XMM-Newton} has been reported to be 
consistent with the CDM picture 
\citep{arabadjis02, lewis03, katayama03, buote04}.
However, the mass profile around the optical body of elliptical galaxies 
has not been constrained, since the previous X-ray observations could not
measure the brightness and temperature profile at the inner region of 
galaxies.
In addition, there is still a lack of information on DM content in
X-ray faint elliptical galaxies.

Recently, several other techniques become available to probe the mass
profile around elliptical galaxies: strong
gravitational lensing \citep{kochanek95} at 1--10$r_e$ of
distant galaxies, and kinematics of stars, nearby dwarfs,
and planetary nebula \citep{kronawitter00, gerhard01, kleyna02, romanowsky03}
at (0.1--5)$r_e$.
These observations revealed that the mass-to-light ratio is almost
constant at 3--30 around with radius of (0.5--5)$r_e$, 
and possibly correlates with the optical luminosity.
Analyses of the Einstein and ROSAT HRI data
derived a mass profile of X-ray bright elliptical gaiaxies down to $\sim r_e$,
and confirmed that the innermost mass profile from X-ray observations is 
consistent with those from other techniques 
\citep{paolillo02, paolillo03, mathews03}.
Presence of DM around elliptical galaxies are also implied by the
difference between X-ray and optical distributions \citep{buote98,
buote02}.

Generally, previous X-ray and lensing measurements determined the total
mass at the outer region of $>3r_e$, where the DM
dominates, while kinematics of objects constrain the mass at the inner
region of $<3r_e$, where the stellar mass dominates.
Therefore, the overall mass profile from (0.1--10)$r_e$ for many elliptical
galaxies is still unknown; we are not aware of 
where the cross point between the DM and stellar mass profiles is.
Previous works reported two different pictures of DM content in
elliptical galaxies;
dominance of DM even at the inner region of NGC 4636 
\citep{loewenstein02} and NGC 720 \citep{buote02}, while
no need for dark matter in several less-luminous elliptical
galaxies \citep{romanowsky03}.
These issues are concerned with formation scenario of elliptical
galaxies; they formed through galaxy merger or intrinsically.
Also it should be solved why X-ray bright and faint elliptical galaxies exist.

Such study is also important, 
since a substantial fraction of elliptical galaxies
are central galaxies of galaxy groups and we must consider a 
massive dark matter associated with galaxy groups, apart from that of
galaxies.
ASCA observation of NGC 1399 found that it has a double structure of
gravitational potential \citep{ikebe96}, and it is implied that 
DM forms two distinct scales.
Such a double structure is not predicted by the CDM structure formation.
However, ASCA could not resolve the innermost region within the optical half
radius of NGC 1399, and thus contribution of stellar mass was not well 
studied. 

Nowadays 
X-ray studies allow us to obtain the mass profile with a wide scale from
0.1 to 10$r_e$, and thus give us information on the overall
DM density profile.
Unprecedentedly good angular resolution of {\it Chandra}
enables us to measure the temperature and brightness of ISM in
elliptical galaxies down to 0.1$r_e$.
In addition, a high signal-to-noise ratio and a wide field of view of {\it
XMM-Newton} allow us to trace the DM profile beyond the optical
outskirts even for X-ray dim elliptical galaxies.
Previously several workers reported on the {\it Chandra} and {\it
XMM-Newton} studies of DM profiles in elliptical galaxies 
\citep{loewenstein02, buote02, osullivan04a, osullivan04b},
but sample is still poor, and 
a general view of DM in elliptical galaxies is not yet known.
Here, we report on studies of 53 elliptical galaxies 
observed with {\it Chandra} and deprojection analyses of 6 elliptical
galaxies observed with both {\it Chandra} and {\it XMM-Newton}, 
to examine mass profiles of elliptical galaxies.
Throughout this paper, we assume $H_0=70$ km s$^{-1}$ Mpc$^{-1}$.

\section{Observation and Results}

\begin{figure}
\plotone{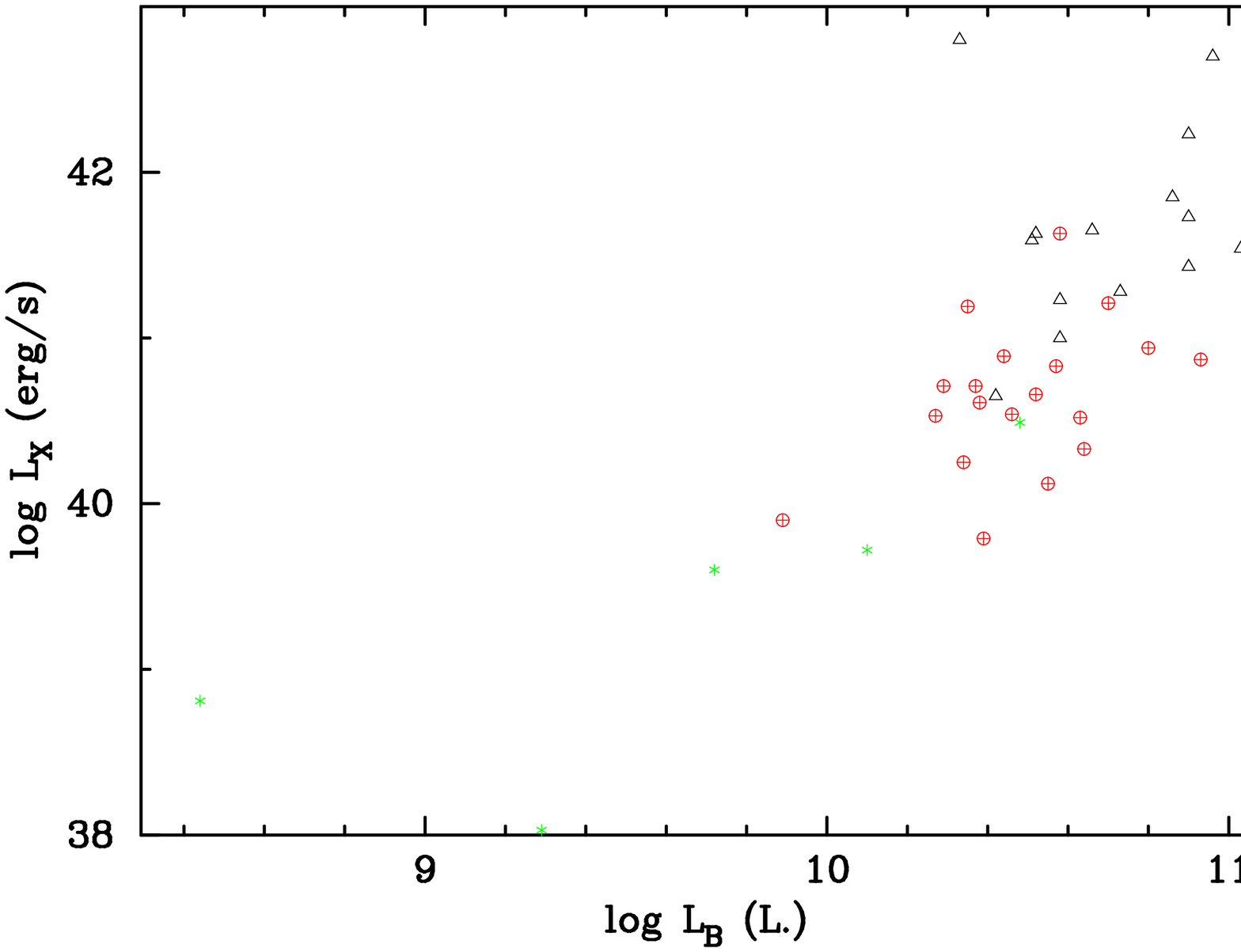}
\caption{The diagram of $L_{\rm X}$ and $L_{\rm B}$ for our sample
 elliptical galaxies. $L_{\rm X}$ is taken from O'Sullivan et al (2001).
Galaxies with triangles, circles, and asterisks are EXG, CXG, and VCXG,
respectively, which are defined in \S\ref{para-ana}.
\label{lblx}}
\end{figure}

We analyzed archival data of 53 elliptical galaxies observed with {\it
Chandra} ACIS instruments during 1999-2003 (table \ref{tab:sample-chandra}).
These galaxies are selected so that their exposure time or the total
photon count is more than 18 ks or 10,000 cts, respectively, 
to achieve a good signal-to-noise ratio. 
Figure \ref{lblx} shows the relation of the optical and 
X-ray luminosity of our sample galaxies.
Their optical luminosity $L_{\rm B}$ is in the range of 
$10^{9.5-11}L_{\odot}$, and
thus they are ordinary or giant elliptical galaxies, while X-ray luminosity
scatters widely from $10^{39.5}$ erg s$^{-1}$ to $10^{42.5}$ erg s$^{-1}$.
All of galaxies lie at a redshift of $<0.033$, and 
most of galaxies were observed with the back-illuminated CCD chip,
ACIS-S3, and 6 galaxies were with the ACIS-I.
NGC 4697 has been observed several times, and a
total exposure of {\it Chandra} data is 145 ks in 4 pointing
observations in 2000 and 2003--2004.
Data analysis is performed with the CIAO software package version 2.3
and the calibration data base version 2.21. 
We reprocessed the level 2 data from the level 1 data, and excluded time
regions with high background rate.
We also eliminated point sources identified with the tool {\it wavedetect}.
We constructed response matrices, considering the source extent by using
the tool {\it acisspec},
and corrected the low energy efficiency with the tool {\it acisabs}.
Background is estimated in the same detector region as on-source data 
by using the archival background data indicated by the tool 
{\it acis\_bkgrnd\_lookup}.
We checked with eye the background level by comparing the light curves, the
spectrum of the whole regions of the main CCD chip, and the X-ray radial
count profile between the on-source and background data.
In most of the data, the background was well reproduced within 20\% 
accuracies, but for some
objects (NGC 720, NGC 1700, NGC 1705, NGC 2434, NGC 4494, and NGC 5253), 
the background level of the on-source data is somewhat higher
due to the continual particle flare.
Since the excess background is harder and the thermal emission from
elliptical galaxies are dominant at 0.6--1.5 keV around the Fe-L line complex, 
it is not so significant for our analysis.

We also utilized the {\it XMM-Newton} archival data of some elliptical
galaxies in table \ref{tab:sample-chandra}, 
to trace the mass profile far away from the optical isophoto.
About half of our sample galaxies were observed with {\it XMM-Newton}.
Since the outer region of X-ray bright galaxies has already been 
investigated well with {\it ROSAT} and {\it ASCA}, 
we here treat two representative X-ray bright elliptical galaxies, 
NGC 1399 and NGC 5044.
Since we intend to investigate the mass profile from the innner to the
outer region, we chose the objects whose data of both {\it Chandra} and
{\it XMM-Newton} have a good signal-to-noise ratio.
Then, 8 X-ray faint galaxies remain, but two objects
(NGC 4552 and NGC 5171) significantly suffer the Galactic soft X-ray 
emission around them, and two objects
are associated with the intragroup medium (NGC 4261 and NGC 7619).
As a result, we analyzed the data of four X-ray faint elliptical galaxies,
NGC 3923, NGC 4697, NGC 720, and IC 1459, around which we do not find
any significant flat X-ray emission.
Observations of six galaxies analyzed here are summarized in table 
\ref{tab:sample-newton}.
We processed and analyzed the EPIC data with {\it XMMXAS} ver. 6.0.0.
We constructed response matrices, considering the source extent.
Time regions with high background rate are excluded.
We also excluded the point source regions in the {\it XMM-Newton} data.
Here point sources were picked up as the same method in
\citet{fukazawa04}, but the exclusion radius is set to 20$''$.
Background is estimated in the same detector region as on-source data 
by using the archival background data (Jan. 2002 version).
For most of the data,
the background level is well reproduced within 30\% and 60\% accuracies
for the MOS and PN, respectively, in the energy range of 0.5--5 keV.
On the other hand, the background level of NGC 1399 is much higher
due to the heavy particle flare.
Fortunately, the X-ray emission of NGC 1399 is relatively bright,
the thermal emission around the Fe-L line complex is still higher than
the background level even at the outer galaxy region, and thus, 
we restricted the energy band for analyses within 0.8--2.0 keV.

\section{Results}

\subsection{Overall Spectral Features}\label{overall}

Before the datailed analysis, we analyze an overall spectrum obtained
with the {\it Chandra} ACIS, to obtain overall spectral properties.
We accumulated the spectra centered on the galaxy nucleus within the
maximum detection radius $R_{\rm max}$, which is described in the
subsequent section.
In order not to use the CCD chip edge, we set the outer radius to
240$''$ when $R_{\rm max}$ is larger than 240$''$.
For some galaxies, we changed the integration radius so that the central
point sources or the ambient background diffuse emission are not
included in the spectrum.
Then we fitted each spectrum with the APEC model \citep{smith01} plus
bremsstrahlung multiplied by the Galactic absorption.
The bremsstrahlung model represents the emission from unresolved point
sources and we fixed the temperature to 7 keV \citep{matsushita94,
matsumoto97}.
The column density of the Galactic absorption is fixed to the value
referred to \citet{dickey90}. 
Some galaxies show an indication of excess absorption, and we left the
column density free for such galaxies.
The results are summarized in table \ref{table:spec}.
For X-ray faint galaxies, the spectra were well fitted with this model, and the
flux of the bremsstrahlung is consistent with the contribution of
unresolved point sources in each galaxy.
Our results of the temperature for NGC 3585 and NGC 4494 is also
consistent with that of \citet{osullivan04a}, who analyzed the {\it
Chandra} and {\it XMM-Newton} simultaneously.
On the other hand, X-ray bright galaxies give a large reduced-$\chi^2$.
This is thought to be due to the significant temperature gradient, as
shown in the next subsection.
Several galaxies give a very high luminosity of the hard component with
$>10^{41}$ erg s$^{-1}$, possibly due to the imcomplete modeling of 
the thermal emission.
Since the hard component is not so significant for our studies, we do
not discuss it furthermore.

\subsection{Parameterization studies on {\it Chandra} data}\label{para-ana}

Temperature and density profiles of ISM are necessary to
derive the mass profile.
The most accurate is to obtain these profiles from analysis of 
deprojected spectra.
However, signal-to-noise ratio of data for many of X-ray dim galaxies
does not allow us to measure the mass at enough number of radii.
Therefore, we first obtained the temperature profile from the projected
spectra at several annuli, and the ISM density profile from the X-ray
surface brightness profile by ignoring the temperature dependence of 
emissivity, and then parameterized these two profiles by
appropriate functions from which we derived the mass profile.
In this method, projection effect on the temperature profile and
neglect of the temperature dependence of emissivity gives an error
to the mass estimation, but it is at most 30--40\% which is smaller than
the statistical error in the deprojection analysis.
Furthermore, as described later, the mass profile derived here is
found to be consistent with that derived by the deprojection analysis.
Therefore, the parameterization studies are valuable to obtain an
overall picture of radial profiles of the temperature, gas density, and
mass for elliptical galaxies.

We divided a galaxy region into several rings so that the 
signal-to-noise ratio is more than 15, which is enough to constrain the
temperature by utilizing center energy of strong Fe-L lines.
The APEC model 
is usually employed for fitting the thermal X-ray emission, but
convergence of fitting is slower than the MEKAL \citep{liedahl95} and
the fitting often falls into the local minimum.
In this analysis, we treat many radially sorted spectra, and thus 
we do not apply the APEC model.
Then we fitted each spectrum with the MEKAL model plus
bremsstrahlung multiplied by the Galactic absorption.
The column density of the Galactic absorption is fixed to the value
referred to \citet{dickey90}.
Allmost all of the spectra were well fitted with this model, and the
flux of the bremsstrahlung is consistent with the contribution of
unresolved point sources in each galaxy.
The metal abundance cannot be determined accurately, and thus we do not
discuss it.
Thus-obtained temperature profile of each galaxy is represented by a
polynomial function with 0, 1, 2, or 3th order.
Since several galaxies show a complex profile, we fitted the profile
at a radius of 0--50$''$ and 50$''$--, separately, with a 2th-order
polynominal, and jointed two polynomials smoothly.
The order of the polynomial function is the same between the inner and
outer region.
For all the galaxies, we obtained good fits to the radial temperature
profile.
In the case that the number of the radially-sorted spectra is less than
5, we assume a constant temperature profile, where the temperature is 
obtained in the previous subsection.

In most of galaxies, the X-ray image is not circularly symmetric: 
the hot gas at the center region often suffers the jet from the 
galactic nucleus \citep{jones02, buote03, kraft04}, while the outer
region is sometimes elongated toward the specific direction.
In the case of the jet disturbance, the assumption of hydrostatic
equilibrium does not completely hold.
Therefore, we must note on the correctness of mass estimation.
However, the assymmetry is as small as those of the X-ray emission of
galaxy clusters, for which it was reported that the assymmetry does not
affect the mass estimation so much \citep{schindler96, katayama03}.
The X-ray surface brightness profile is derived at an energy of 0.5--1.5
keV, and fitted with single, double, or triple $\beta$ model.
Profiles of most of the galaxies could be fitted well by either of the
above three $\beta$ models with $<$10\% accuracies.
In several galaxies, the radial X-ray surface brightness profile has
local structures, but the deviation is at most 20\%, which gives 10\%
error to the hot gas density.
Point-like emission from the galactic nuclei is observed for NGC 1553,
NGC 4261, and IC 1459, and we modeled its X-ray profile by the
$\beta$-model with $\beta=0.83$ and core radius of 1.68$''$, which are
obtained by fitting the X-ray profile of the point source M33 X-8.

In table \ref{table:results}, 
we summarized the results of analysis on temperature and 
surface brightness profile for sample galaxies.
The temperatures in this work are almost consistent with those obtained
by the ROSAT PSPC \citep{matsushita01} for 29 galaxies, but the
temperature at the center region, $T_{\rm i}$ in table
\ref{table:results} is sometimes lower for galaxies with a strong
positive temperature gradient, due to finer angular resolution 
of {\it Chandra}.

\begin{figure}
\plottwo{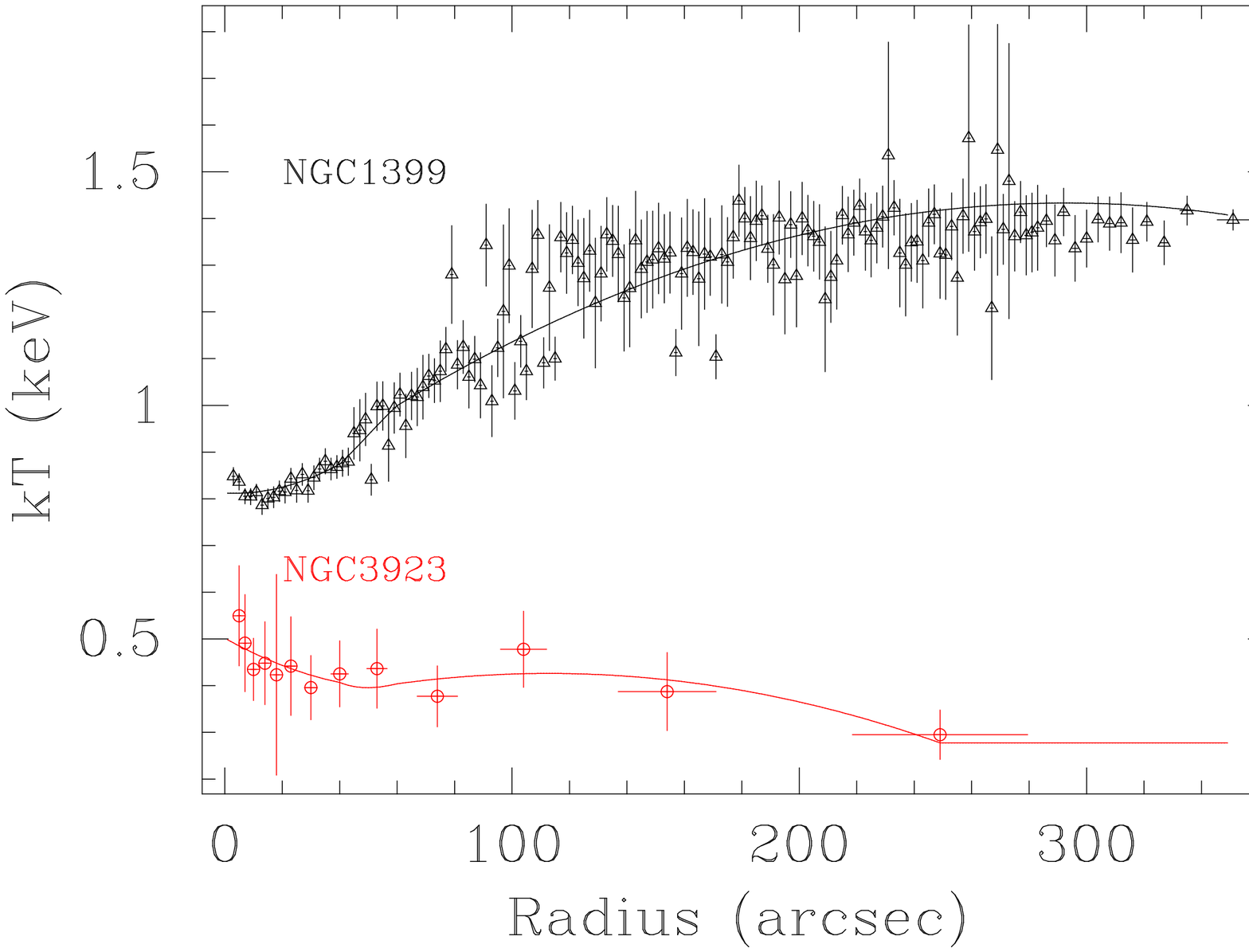}{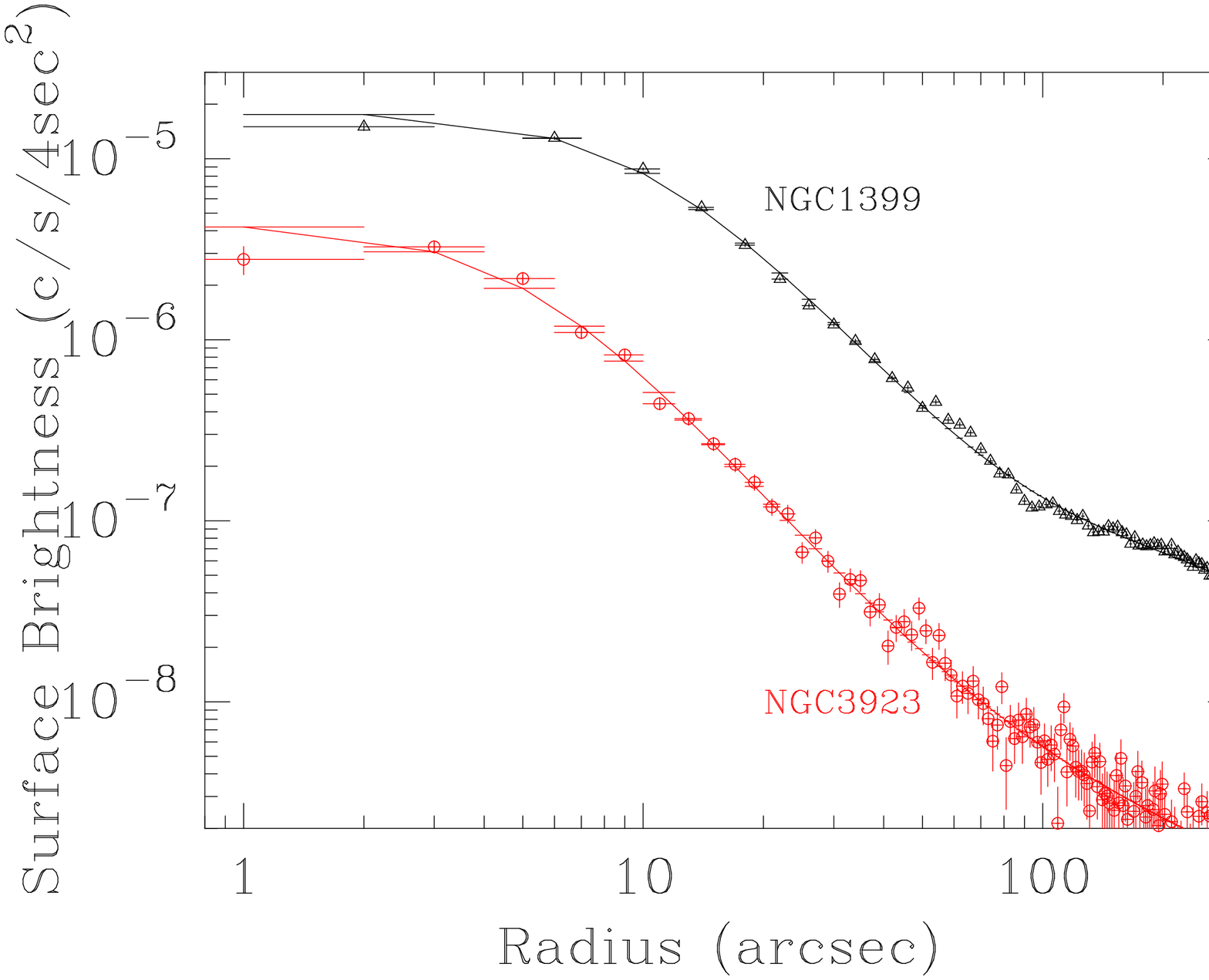}
\caption{The left and right panels show profiles of temperature and
 X-ray surface brightness (0.5--1.5 keV) of NGC 1399 and NGC 3923. The
 solid lines represent the best-fit parameterized function. see the text
 in detail.
\label{r-ktsb}}
\end{figure}

\begin{figure}
\plottwo{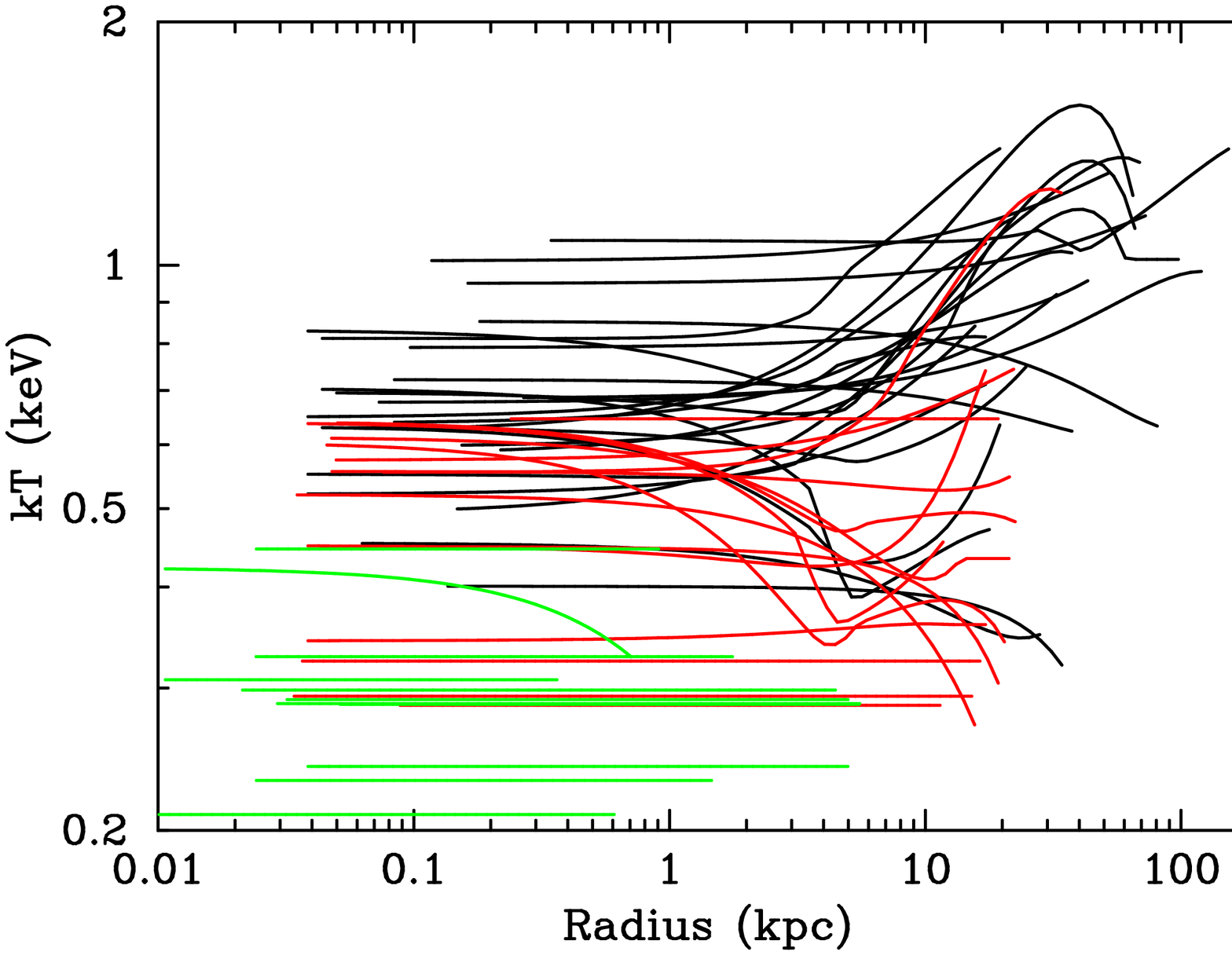}{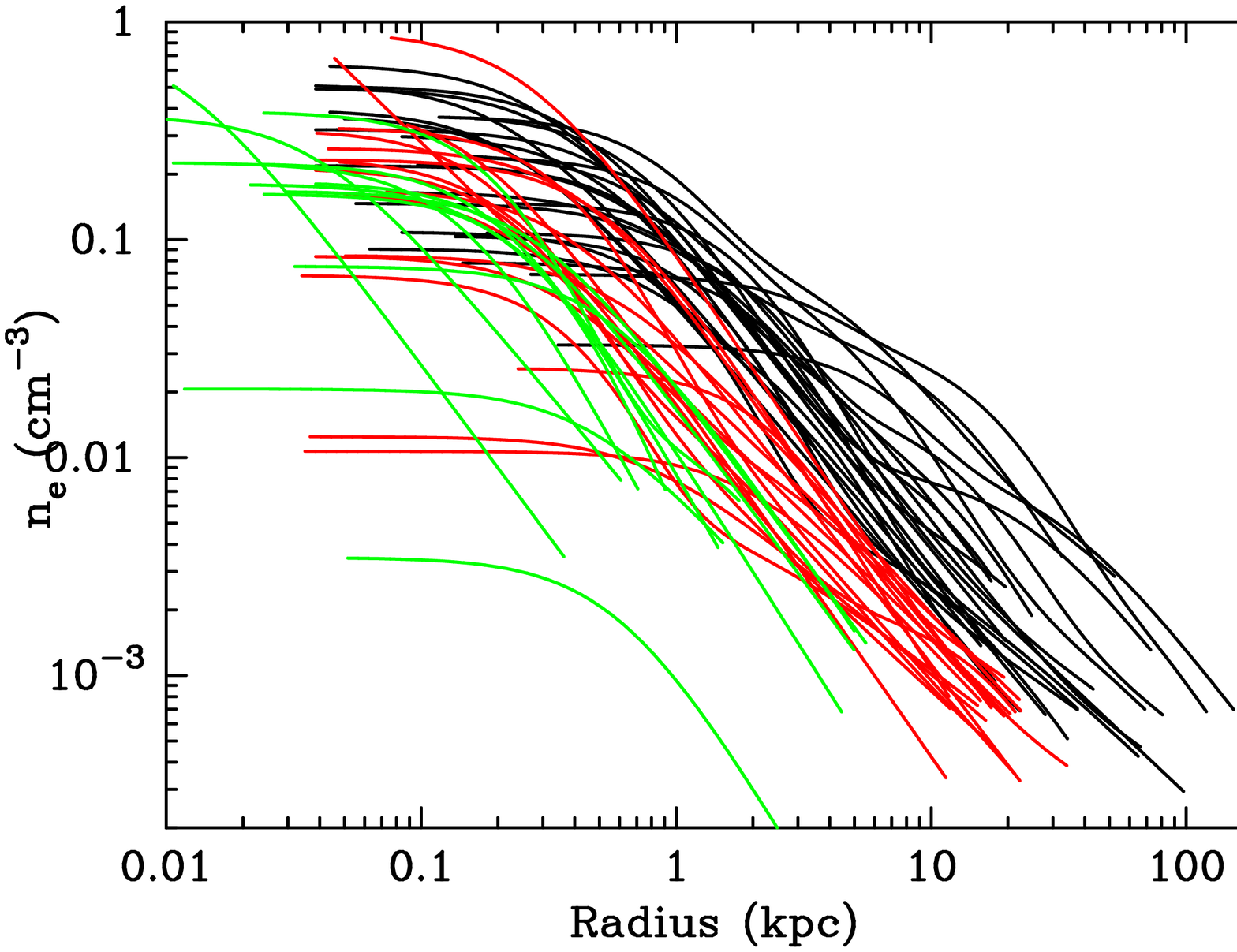}
\caption{The left and right panels show the parameterized radial profiles of 
temperature and gas density of sample galaxies. 
The black, red, and green lines represent the EXGs, CXGs, and VCXGs,
respectively.
\label{r-ktne}}
\end{figure}

Examples of two typical profiles of temperature and X-ray surface
brightness are shown in figure \ref{r-ktsb}, 
together with the parameterized function.
Left panel displays those of a X-ray bright galaxy NGC 1399, wherein the
temperature gradually increases toward the outer radius from 0.8 to 1.4
keV, and the X-ray emission is detected beyond the ACIS-S field of view.
In contrast, right panel of a X-ray dim galaxy NGC 3923 presents a
flat or declining temperature profile at 0.4--0.5 keV and a compact
X-ray emission region for the same surface brightness. 
In figure \ref{r-ktne}, 
we plot the temperature and ISM density profiles of objects 
analyzed here, where the temperature profile is plotted for objects
with more than three radial bins.
These profiles are plotted up to the maximum radius $R_{\rm max}$, within 
which the source brightness is higher than 
$1\times10^{-9}$ cts s$^{-1}$ pix$^{-2}$ cm$^{-2}$.
When the X-ray emission from the object is extended beyond the {\it
Chandra} field of view, we plot the profile within $240''$.
Around some galaxies, 
there is a significant flat emission at the outer region, which
is due to the excess instrumental background, the bright Galactic X-ray 
background, or the Virgo cluster X-ray emission. 
In that case, we set the $R_{\rm max}$ to be small.
We divide our sample into two galaxies, based on whether the ISM density at 10
kpc radius is higher or lower than $2\times10^{-3}$ cm$^{-3}$.
Hereafter we call them an extended X-ray galaxy (EXG) or a compact
X-ray galaxy (CXG), respectively.
The classification is shown in table \ref{table:results}.
In addition, galaxies whose X-ray emission does not reach 10 kpc are
classified into a very compact X-ray galaxy (VCXG), 
because we can obtain the ISM properties of such objects to some extent but 
cannot constrain their total mass profile
to say anything about the DM profile.

EXG has a higher ISM density by 3--10 times than CXG at the outer
region, and tends to show a positive gradient in the temperature
profile, like NGC 1399.
Looking at the X-ray images of ROSAT and ASCA, 
the X-ray emission of some EXGs is extended beyond 100 kpc.
The temperature at the outer region is 0.9--1.3 keV, corresponding to
that of galaxy groups.
These phenomena indicate that the EXG is embedded by the ambient
hot intragroup medium.
In fact, several EXG galaxies are identified as a cD galaxy in the
galaxy group, such as NGC 1399, NGC 5044, and NGC 4325.
On the other hand, CXG and VCXG show a flat or declining temperature 
profile like NGC 3923, 
and the X-ray emission is detected up to at most several tens kpc.
Therefore, no significant hot intragroup medium is seen
around CXG and VCXG.
The central temperature ranges from 0.2 to 1.0 keV, and the temperature
of all EXGs is higher than 0.5 keV.
As seen in figure \ref{lblx}, EXG and CXG+VCXG galaxies correspond to the
X-ray bright and faint ellipticals, respectively.
However, this classification does not strictly mean that EXG or
CXG is surrounded by the intragroup X-ray emission or not, since some
galaxies associated with the intragroup X-ray emission are included in
CXGs: for example, NGC 4261 \citep{davis95} and NGC 7619 \citep{fukazawa04}.

Assuming the hydrostatic equilibrium, we derived the total mass profile
$M(r)$ from the parameterized temperature $T(r)$ and gas density $n(r)$
profiles, following the equation,
\[
 M_{\rm total}(<r)=-\frac{kT(r)r}{\mu m_pG}\left(\frac{d\ln{n(r)}}{d\ln{r}}+\frac{d\ln{T(r)}}{d\ln{r}}\right).
\]
The total mass is obtained for several elliptical galaxies in our
sample, by the
optical measurements of kinematics of stars, planatory nebulas, and
dwarf galaxies.
We compare the total mass $M$ at a given
radius, to that obtained by the previous optical works.
In most cases, not the total mass $M$ but the mass-to-light ratio
$M/L_{\rm B}$ is available.
To derive the mass-to-light ratio, the light profile $L_{\rm B}(r)$ is
nesessary.
We assumes the de Vaucouleurs law for the 
stellar distribution, while optical workers derived it from the
optical observation.
This difference possibly gives a different mass-to-light ratio.
Therefore, when the total mass $M$ is available, we compare it with our
result.
Since \citet{kronawitter00} summarizes many data of $M/L_{\rm B}$, we
convert their value of $M/L_{\rm B}$ at their optical-half radius to the
total mass $M$, by using half of the total optical luminosity used in
the literature.
Since the $L_{\rm B}$ at the optical-half radius is not explicitly
written in the literature, this conversion may associate the systematic error.
Furthermore, $M$ depends on the distance to the source, and thus 
we corrected the reference values, by using the distance used in this work.
The comparison is shown in table \ref{table:nfwcmp}.
Our values are almost consistent with the optical works within 30\%
errors, but three galaxies show a large difference by a factor or $\sim2$.
This is partly due to the systematic error in converting the 
$M/L_{\rm B}$ into the total mass, for NGC 1399 and NGC 5846.
Quite a large $M/L_{\rm B}$ for NGC 3379 in our values indicate that
the ISM is not in the hydrostatic equilibrium or the contribution from
unresolved point sources to the X-ray emission is significant.
Our result for NGC 720 and NGC 4555 is almost consistent with those 
in \citet{buote03} and \citet{osullivan04b}, respectively,
who analyzed the same X-ray data as ours.
In addition, our values are consistent with those estimated 
by the stellar population synthesis \citep{gerhard01}.
In figure \ref{n4697mass} top, we compare the mass profile in this work
with that in the optical works \citep{kronawitter00, saglia00}, 
for NGC 4472 and NGC 1399.
The mass profile is consistent with each other, and our mass profile of
NGC 1399 beyond 10 kpc is also consistent with that 
in the previous X-ray measurements with ASCA \citep{ikebe96}.
These demonstrate that the X-ray mass measurement around $r_e$ is consistent
with that in the optical works, indicating that the hot ISM is roughly in
the hydrostatic equilibrium.
Accordingly, we can trace the mass profile in a wide range of $(0.1-10)r_e$.

\begin{figure}
\plottwo{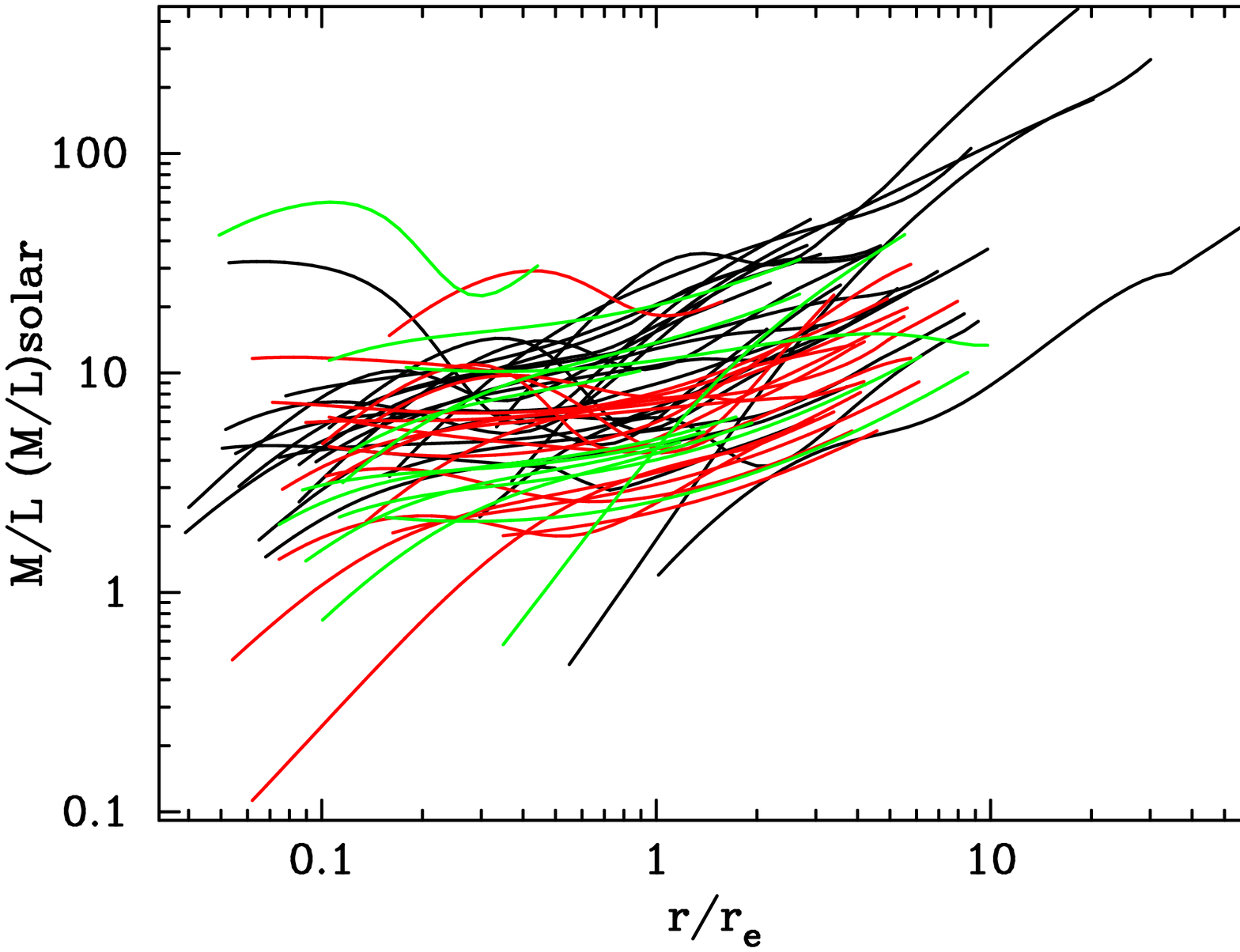}{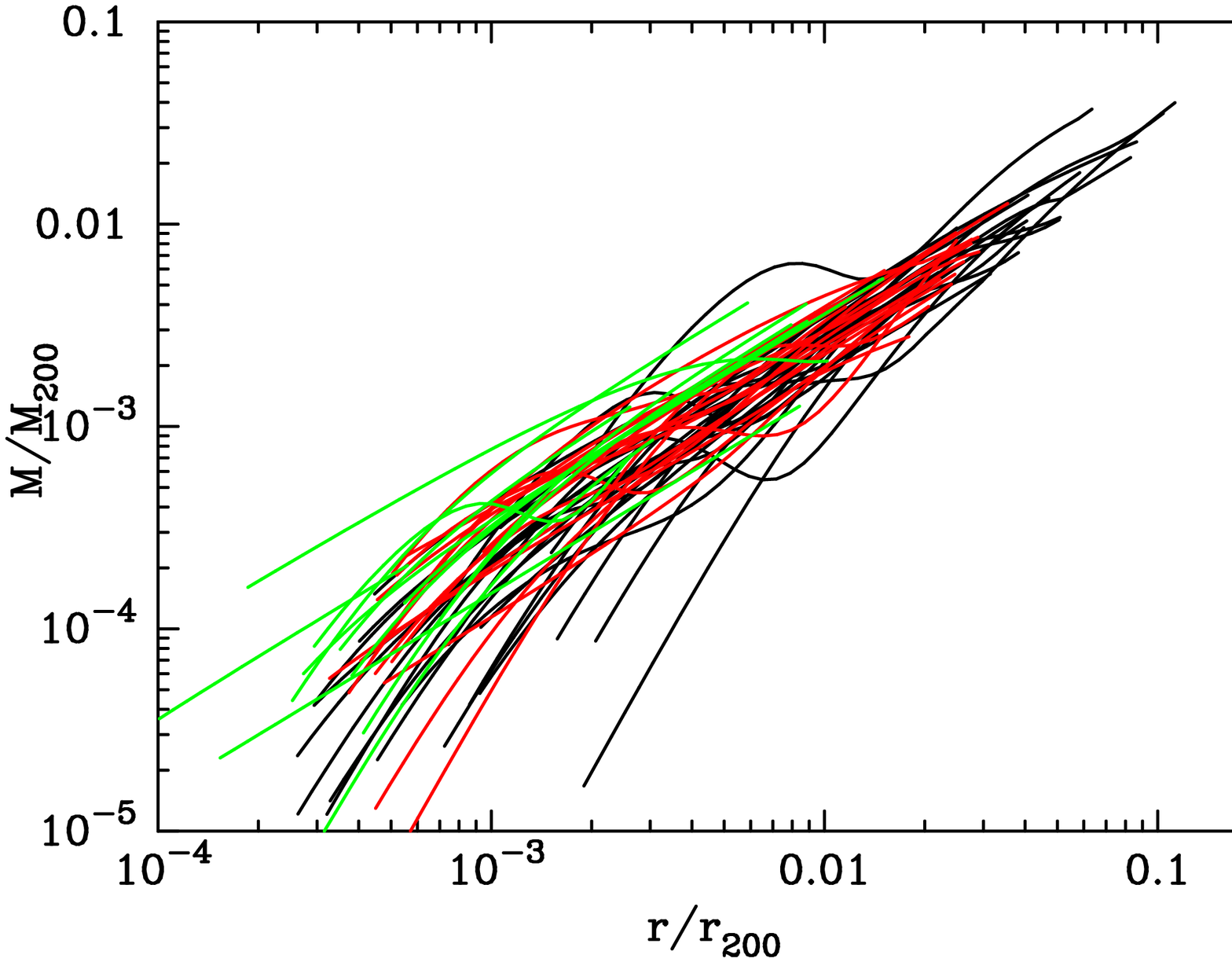}
\caption{The left and right panels show the profiles of
 mass-to-light ratio and scaled total mass,
 respectively, of sample elliptical galaxies. Line colors are the same
 as those in figure \ref{r-ktsb}.
\label{r-mls}}
\end{figure}

We then derive the profile of
mass-to-light ratio $M(r)/L_{\rm B}(r)$, scaled total mass
$M(r)/M_{200}$, where $r$ is a radius.
A scaling virial mass $M_{200}$ is calculated as 
$\frac{4}{3}\pi\left(200\rho_{\rm crit}\right)r_{200}^3$, 
referring to \citet{evrard96}.
We scale $M$ profile with a scaling (virial) radius
$r_{200}=1167\left(kT_{\rm o}/{\rm 1 keV}\right)^{0.5}$ kpc, and $M/L_{\rm B}$ 
profile with an optical-half radius $r_e$.
Here, $kT_{\rm o}$ is a temperature at 10 kpc from the galaxy center
where the dark matter in general dominates, and this temperature is 
calculated from the model of the temperature profile.
The reason why we do not derive the virial radius by extrapolating the
mass profile is that the maximum detection radius is smaller by a
factor of $>10$ than the virial radius for most galaxies, and thus the
extrapolation introduces large uncertainties.
We present the above profiles from $4''$ to the maximum
X-ray detection radius in figure \ref{r-mls},
since a bin number of the radial surface brightness profile is not enough
within a central 4$''$ and thus the modeling is incomplete.

It is clearly seen that, in many galaxies, there is a region where 
$M/L_{\rm B}$ becomes constant around 3--10 $(M_{\odot}/L_{\odot})$.
This region ($<r_e$) corresponds to that where stellar mass dominates.
$M/L_{\rm B}$ values of EXG at this region tend to be larger than that of CXG;
7--25 $(M_{\odot}/L_{\odot})$ for EXG, and 3--7 $(M_{\odot}/L_{\odot})$ for CXG.
On the other hand, $M/L_{\rm B}$ profile has a break around $r_e$, and
$M/L$ gradually rises up toward the outer region.
These indicate that $M/L_{\rm B}$ is not constant around galaxies and thus 
the presence of dark matter is needed.
In comparison to CXG, EXG has a clear break in the profile 
and a breaking radius is small.
The drop of $M/L_{\rm B}$ at the innermost radius is seen in several galaxies.
Since the inner region often suffers the radio jet from 
the central active galactic nucleus (AGN), the ISM is thought to be deviated 
from the hydrostatic equilibrium in those galaxies.
In contrast to a large scatter of the $M/L_{\rm B}$ profile in the outer
DM-dominated region, the scaled mass
profile has a small scatter.
These indicate that the temperature at the outer region well represents
the gravitational potential depth, and the optical-half radius $r_e$ is
not appropriate for scaling the total mass profile.
The slope of the scaling mass profile has a break around 0.01$r_{200}$,
within which stars dominate dark matter in mass.

\subsection{Deprojection Analyses on {\it Chandra} data}\label{chandra-dep}

In the parameterization analyses, we must pay attention to model
dependencies of mass profiles, by considering uncertainties associated
with the parameterizing model functions.
Alternatively, we here performed the deprojection analyses on the {\it
Chandra} data.
We extracted several projected spectra of concentric annulus centered 
on the galaxy.
We chose the radial bins so that the signal-to-noise ratio is more than
40 or 30 for EXGs or CXGs, respectively.
However, as seen in table \ref{table:results},
photon statistics for 40\% of galaxies, most of which are CXGs or VCXGs, 
is too poor to perform the deprojection.
Deprojected spectra were derived from these projected spectra, 
following the method described in \citet{ikebe04}.
Here, we modified their method by considering the emission out of the
deprojected spherical region.
We fitted the X-ray radial count rate profile with
a single $\beta$ model, and utilized the best-fit parameters to
represent the emission out of the deprojected spherical region and
calculate the projection contribution of that region to the inner region.

\begin{figure}
\plottwo{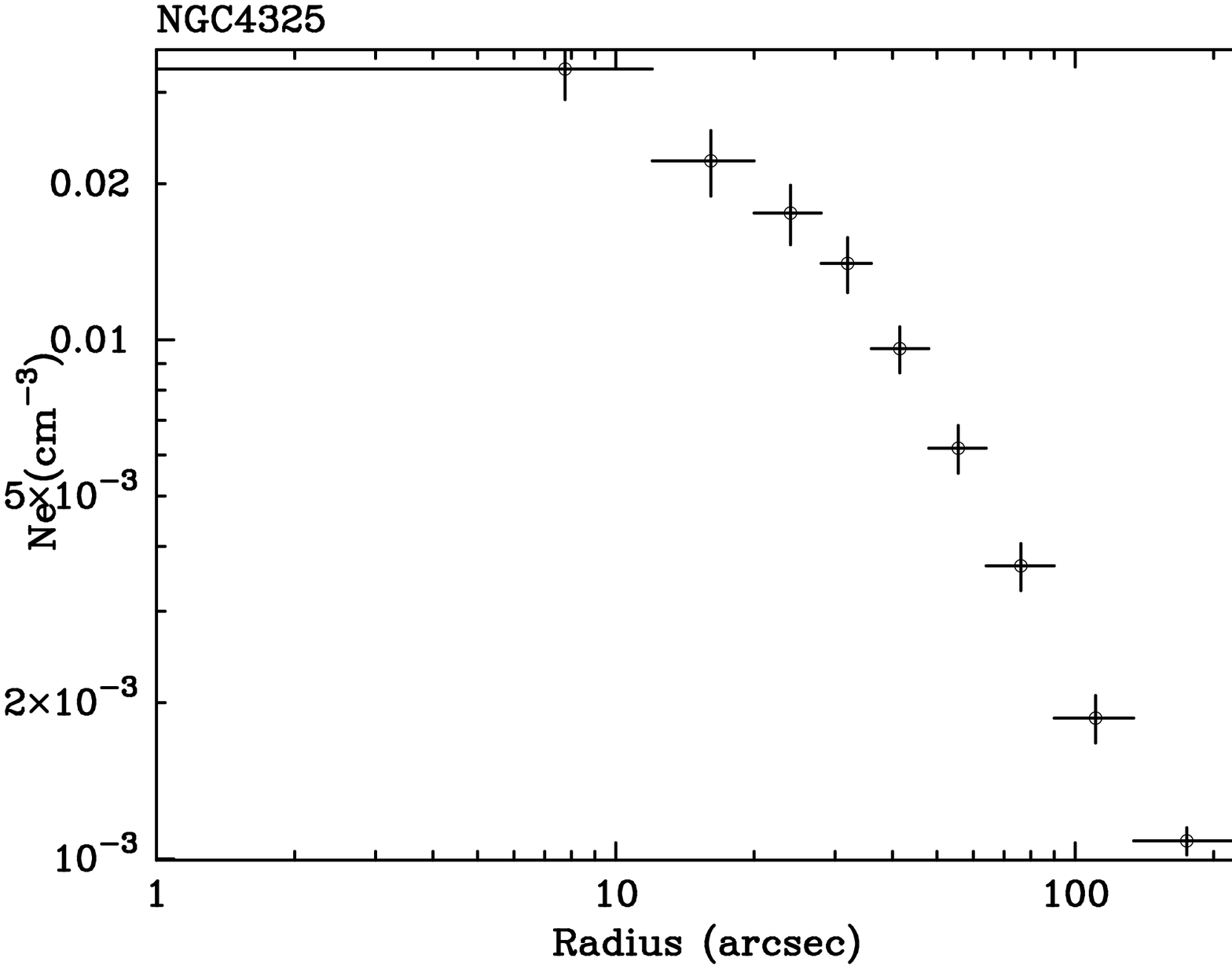}{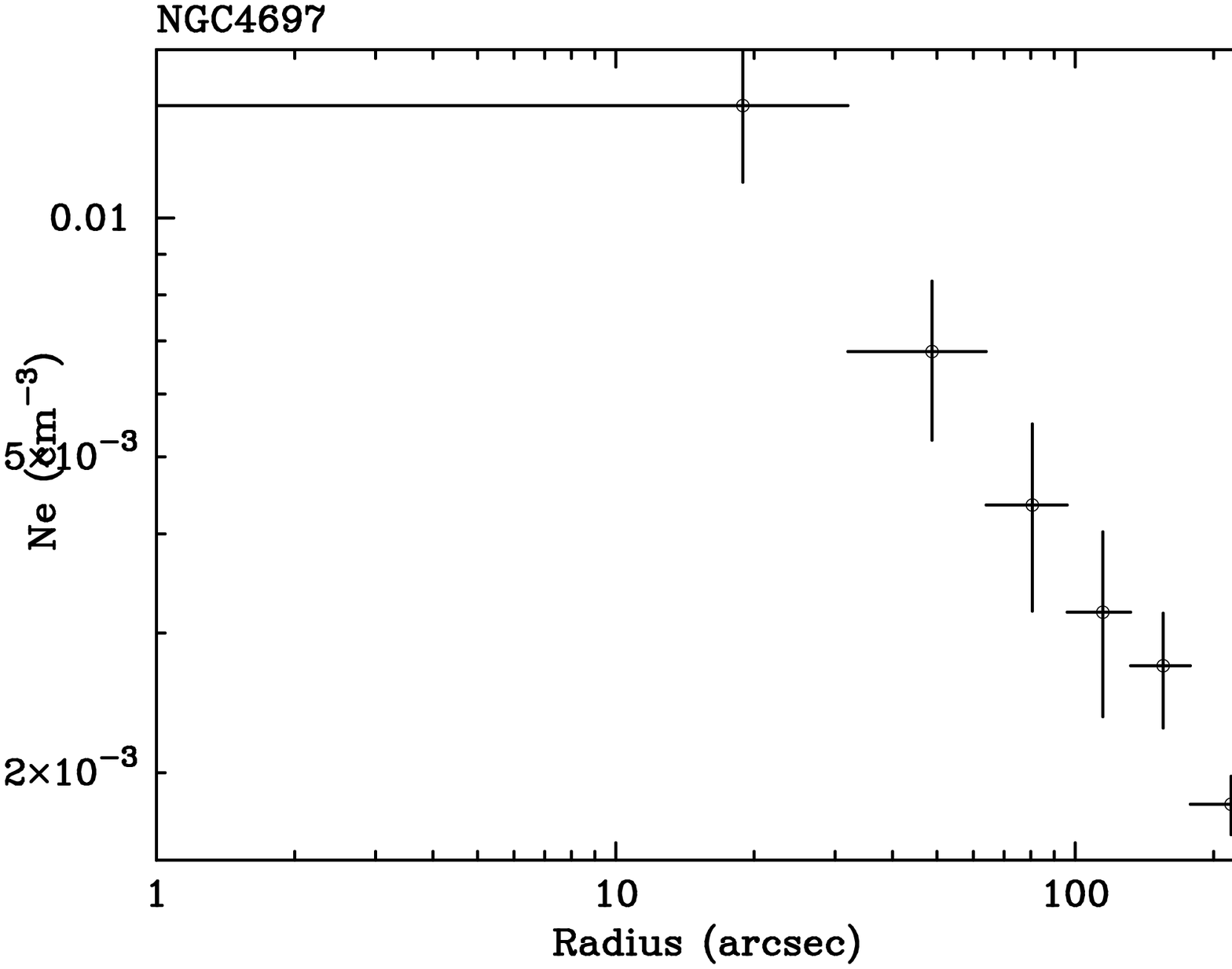}
\vspace*{1cm}
\caption{An example of the gas density profile derived from the
 deprojected spectra. The left and right panels are NGC 4325 and NGC
 4697, respectively.
\label{depn}}
\end{figure}

Deprojected spectra were fitted with one temperature MEKAL plasma model 
multiplied by the photoelectric absorption.
We fixed the absorption column density to the Galactic
value.
Since release of an abundance parameter introduces a large error on the
emission measure, we first fitted the spectra with the abundance free and
took an average of the abundance values over deprojected spectra.
Then, we fixed the abundance to the average value.
For X-ray faint galaxies, we fixed the temperature to the value
which is calculated from the parameterized model function of the temperature
profile obtained in the previous subsection for the corresponding radius.
From the fits of deprojected spectra, we can obtain the 3-dimensional 
temperature and ISM density profiles, $T(r)$ and $n(r)$, where
we took a bin radius as a count-rate-weighted average of radii of all
pixels in the analyzed region.
In figure \ref{depn}, we plot an example of 3-dimensional $n(r)$
profiles obtained here.
We then obtained the mass profiles from $T(r)$ and $n(r)$, 
based on the hydrostatic equilibrium
in which we took a finite difference approximation for calculating a derivative
term.

\begin{figure}
\plottwo{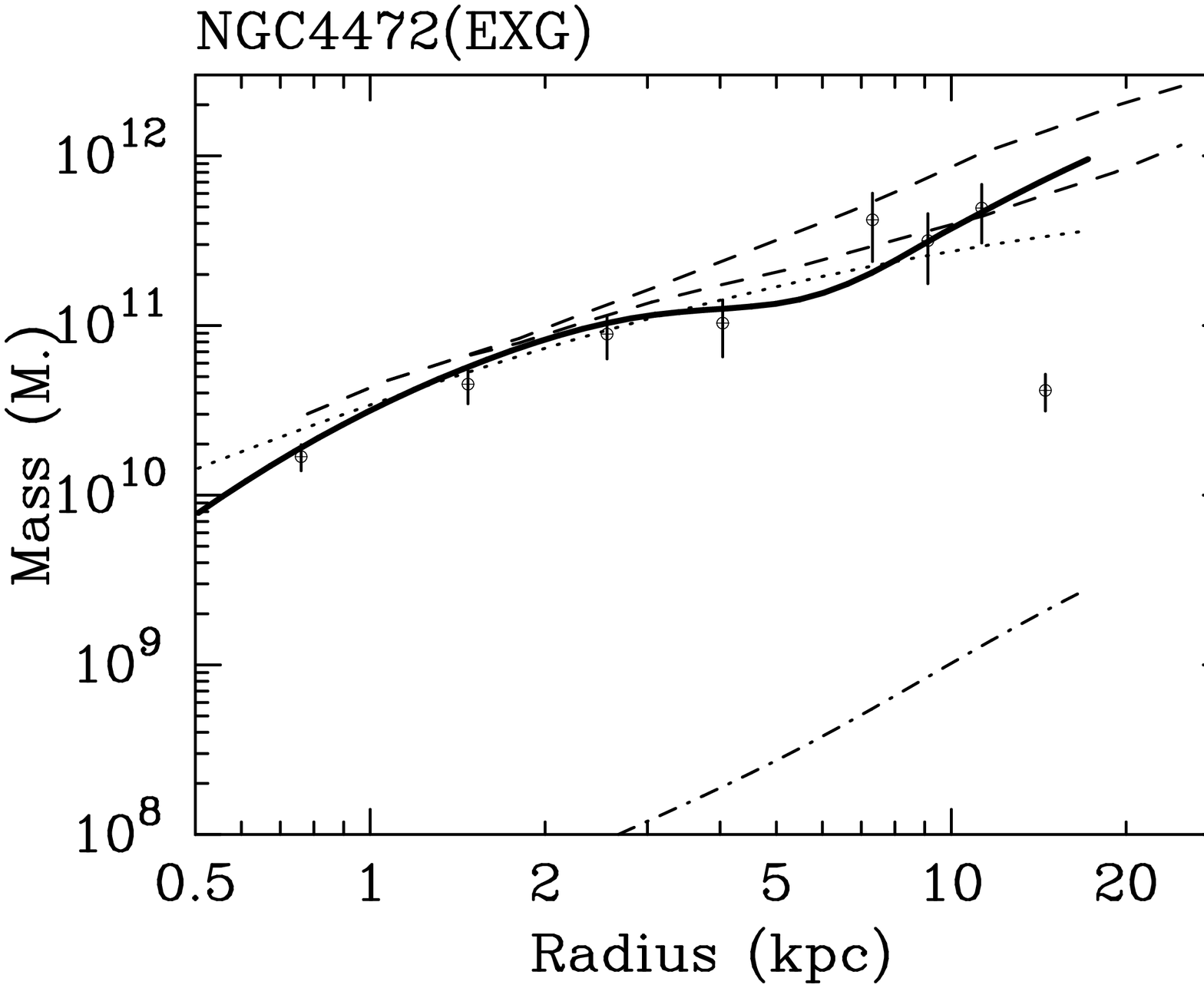}{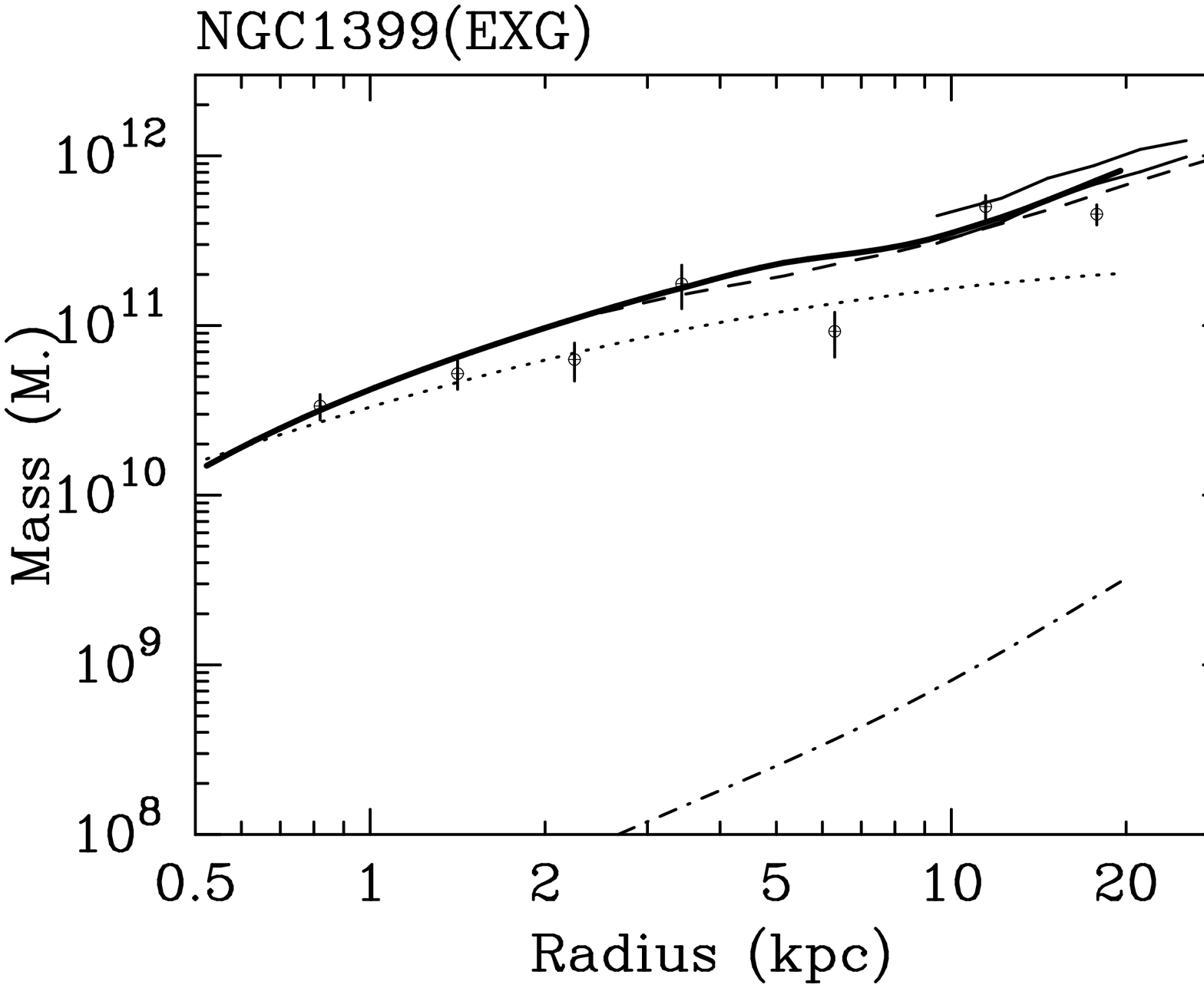}
\plottwo{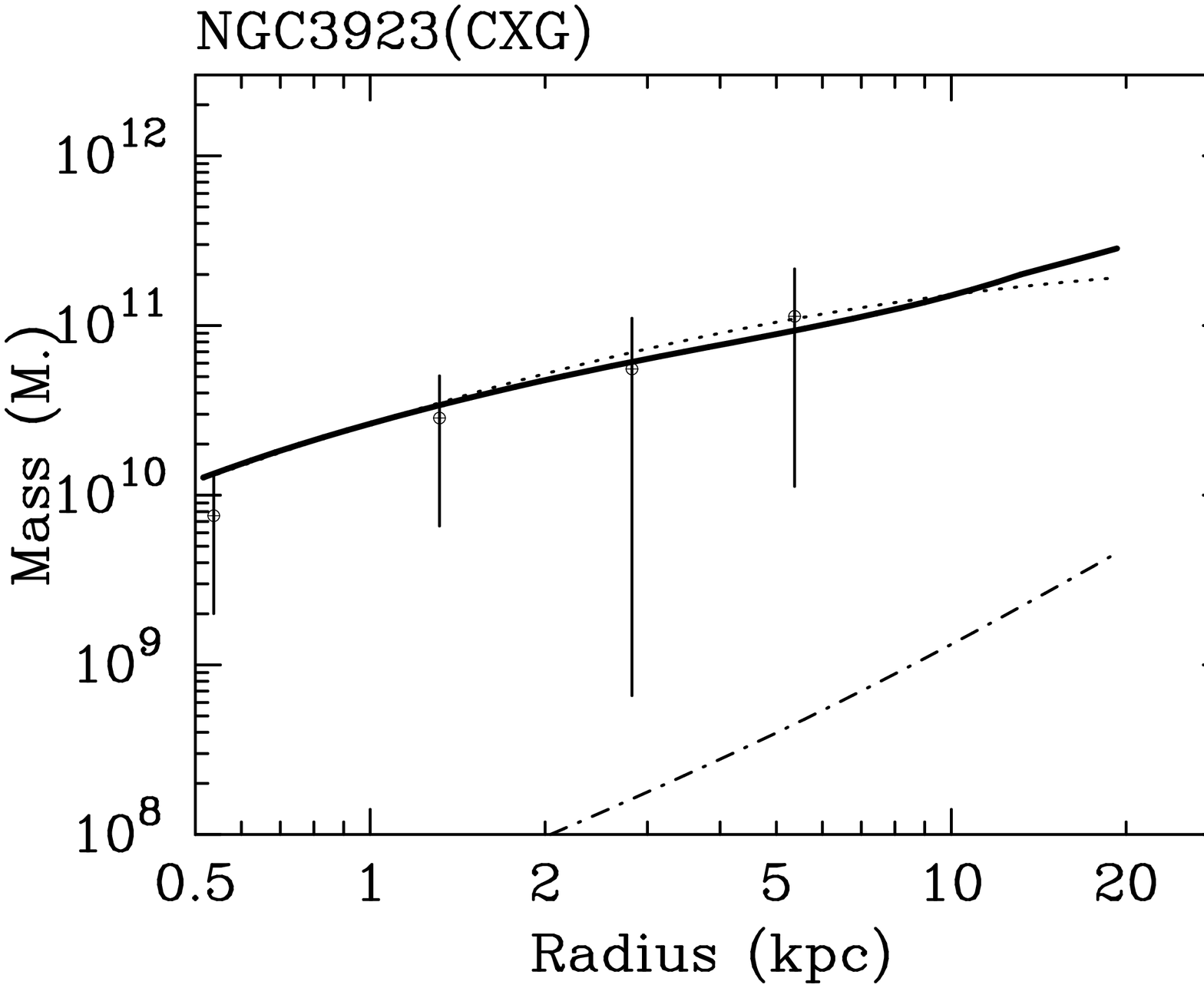}{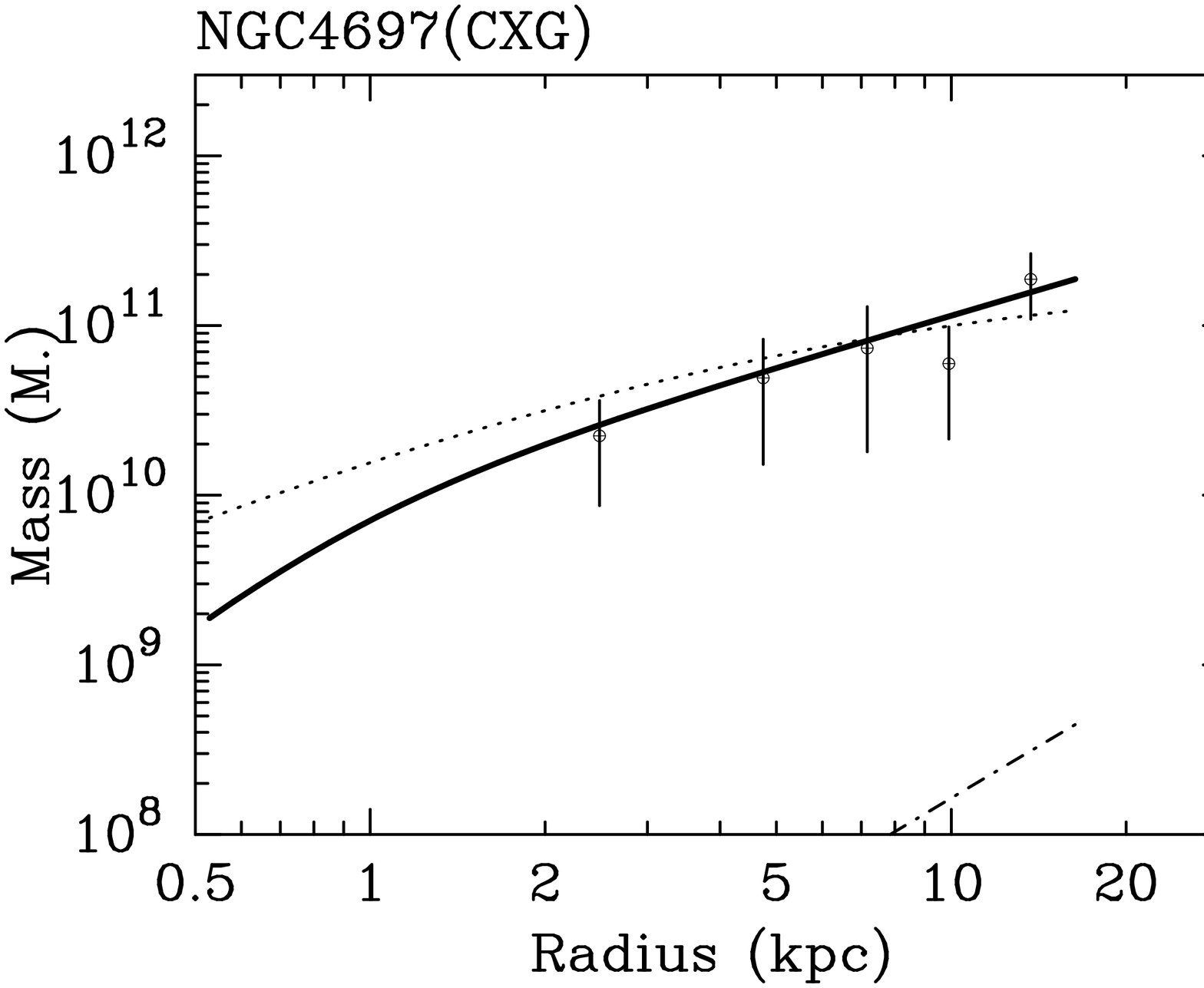}
\epsscale{1}
\vspace*{1cm}
\caption{An example of the mass profiles.
The top-left, top-right, bottom-left, and bottom-right panels are 
NGC 4472 (EXG), NGC 1399 (EXG), NGC 3923 (CXG), and NGC 4697 (CXG), 
respectively.
Data points are the total mass profile obtained by analysis 
of {\it Chandra} data with the deprojection method.
The thick-solid, dot, and dashed-dot line represent the total mass, stellar
 mass, and gas mass, respectively, where the total and gas mass profile
 are obtained by the parameterized analysis and the stellar mass is
 estimated by assuming $M/L_{\rm B}=6(M_{\odot}/L_{\odot})$ for NGC 4472,
 NGC 1399, and NGC 3923 and $4(M_{\odot}/L_{\odot})$ for NGC 4697.
 The dashed lines in NGC 4472 and NGC 1399 are total mass profiles 
 obtained by the
 optical work (Kronawitter et al .2000; Saglia et al. 2000), where two
 lines for NGC 4472 show the permitted mass region.
 Two thin-solid lines represent the
 acceptable range of the total mass obtain with ASCA (Ikebe et al. 1996).
\label{n4697mass}}
\end{figure}

In figure \ref{n4697mass}, 
we plot an example of the mass profile obtained by the
deprojection analysis. 
The total mass profile obtained in the parameterization analysis,
the stellar, and ISM mass profiles are also plotted.
It can be seen that total mass profiles in both analyses are 
consistent with each other.
For NGC 4697, we can measure the mass up to $\sim$20 kpc, thanks to
its long exposure, but it is impossible in most CXGs like NGC 3923.
The total mass seems to be lower than the stellar mass in some galaxies, but
this is thought to be attributed to the incorrect distance $D$ or the
incorrect mass-to-light ratio for the stellar mass. 
Note that the total or stellar mass dependes on the distance 
with $D^{-1}$ or $D^{-2}$.
For example, $D$ to NGC 4697 is assumed to be 15.1 Mpc in this work, and
if we use $D=10.5$ Mpc in \citet{mendez01}, 
the total mass becomes higher than the stellar mass.

\begin{figure}
\plotone{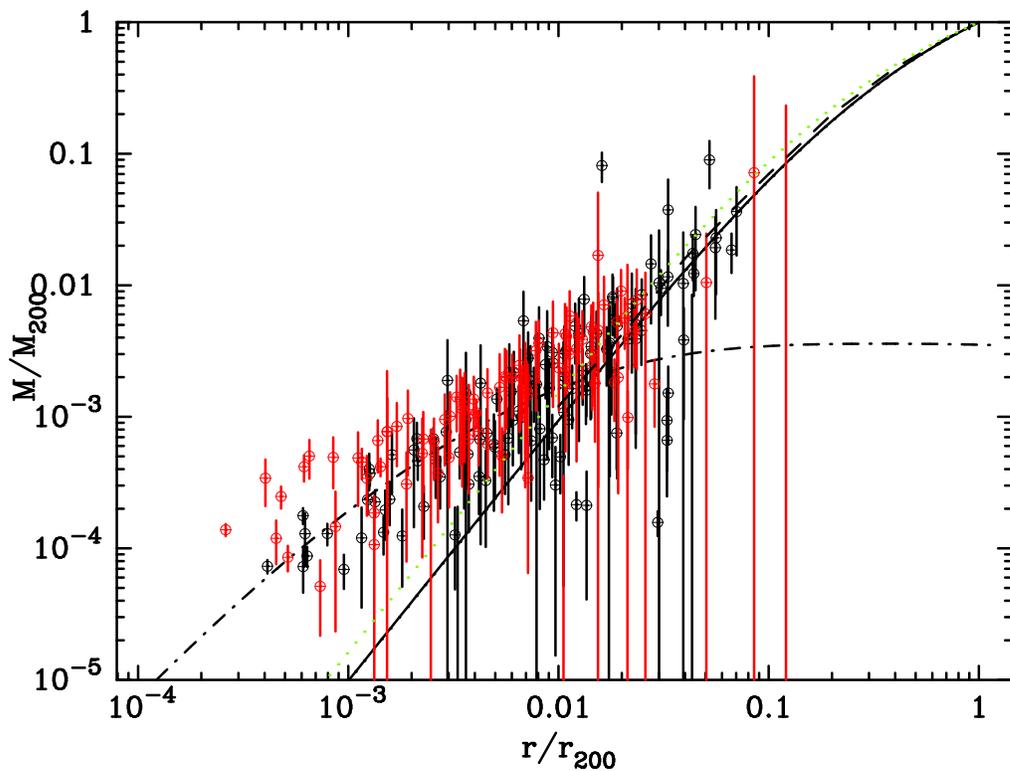}
\caption{Scaled mass profile for sample galaxies
obtained by analysis of {\it Chandra} data with the deprojection method.
Galaxies with black and red points are EXG and CXG defined in \S3.1.
The solid and dot lines represent the scaled NFW profile with $c=4$ and
 $c=6$, respectively, and the dot-dashed line represents a typical 
stellar mass profile, assuming $r_e=8$ kpc, 
$L_{\rm B}=10^{10.5}L_{\odot}$, $M/L_{\rm B}=8(M_{\odot}/L_{\odot})$, 
and virial temperature of 0.6 keV. The dashed line represents the
averaged NFW mass profile for galaxy clusters (Pointecouteau et
 al. 2005). 
\label{nfwcmp}}
\end{figure}

In figure \ref{nfwcmp}, 
the scaled mass is plotted for all the deprojected points of
sample galaxies.
The mass profiles are obtained at (0.001--0.1)$r_{200}$ for EXG galaxies, and
have a concave shape around 0.01$r_{200}$, in good agreement with the
previous subsection.
These data can be compared with the scaled NFW mass profile and stellar
mass profile.
Here, the scaled NFW mass profile is expressed as 
\[
M/M_{200}(x)= \frac{\ln\left(1+cx\right)-cx/\left(1+cx\right)}{\ln\left(1+c\right)-c/\left(1+c\right)}
\]
where $x=r/r_{200}$; here we assume the scaling radius $r_s=r_{200}$.
Therefore, only one parameter $c$ determines the shape of the mass
profile: 
for the same $c$, the scaled mass profile is identical independently of
the system scale.
We show the NFW profile with $c=4$, which is a typical value for rich clusters
\citep{lewis03, buote04}.
The stellar mass profile is different among galaxies, since the effective
radius and virial radius are different.
Here, we plot the typical stellar mass profile, assuming $r_e=8$ kpc, 
$L_{\rm B}=10^{10.5}L_{\odot}$, $M/L_{\rm B}=8(M_{\odot}/L_{\odot})$, and virial
temperature of 0.6 keV.
At the inner region, the mass profile goes along the typical stellar
mass profile, and, 
at the outer region, it well matches the NFW profile.
Comparing with the averaged mass profile of rich clusters in 
\citet{pointecouteau05} (dashed line in figure \ref{nfwcmp}), it is
found that the scaled mass profile almost follows the NFW profile 
from X-ray bright elliptical galaxies (galaxy groups) to rich clusters.
Using the data points beyond 10 kpc where the dark matter dominates, 
we fitted the profile with the
NFW model by varying the concentration parameter $c$, and obtained 
$c=3.61\pm0.75$.
Since the obtained data points can be fitted with the powerlaw
model, and 
the inner slope of mass profile is also fitted to be $1.33\pm0.33$.
This value corresponds to the density inner slope of $\alpha=-1.67\pm0.33$,
somewhat larger than those of rich clusters \citep{lewis03, buote04}.
This is partly due to that the mass used here includes not only dark
matter but also a significant fraction of stellar mass at the inner region.
Most of the above data points come from EXGs, while
data points of CXGs are limited at the inner region of
$<0.02r_{200}$ or $<$(2--3)$r_e$, where the total mass is almost 
explained by the stellar mass.
Nevertheless, we do not see any difference between EXG and CXG, and
there is a hint of the dark matter halo around CXGs in figure \ref{n4697mass}.

\subsection{Deprojection Analyses of {\it Chandra} and {\it XMM-Newton}
data on Six Galaxies}\label{cha-new-dep}

Narrow field of view and low signal-to-noise ratio 
of the {\it Chandra} data do not allow us to investigate the mass
profile up to 3$r_e$, and thus we cannot constrain the dark
matter content for X-ray faint galaxies (CXGs).
Therefore, we utilized {\it XMM-Newton} data to overcome this
limitations.
Here, we analyzed the data of six galaxies, NGC
1399, NGC 5044, NGC 3923, NGC 4697, NGC 720, and IC 1459, which
were observed with both {\it Chandra} and {\it XMM-Newton}.
Combining both data, we trace the mass profiles precisely from inner 
to outer parts of galaxies by deprojection analysis, 
independently of the modeling function.
The procedures of deprojections and spectral fittings are the same as
described in the previous subsection, but in this case, we do not fix
the temperature.
The radii of annulus are not the same between {\it Chandra} and {\it
XMM-Newton} data.
The galaxy center region is finely divided for the {\it Chandra} data, while
the {\it XMM-Newton} data cover the outer region with large steps of 
shell radii.
We took common radii of annulus among three EPIC instruments.
Spectra of MOS1 and MOS2 are summed up before deprojection, and 
we analyzed the deprojected spectra of PN and MOS1+2 simultaneously.
The PN data of NGC 1399 and NGC 4697 cannot be analyzed, due to data
acquisition error or severe background rate.
The obtained profiles of the temperature and ISM density are consistent
between {\it Chandra} and {\it XMM-Newton} within 15\%.

\begin{figure}
\epsscale{.70}
\plottwo{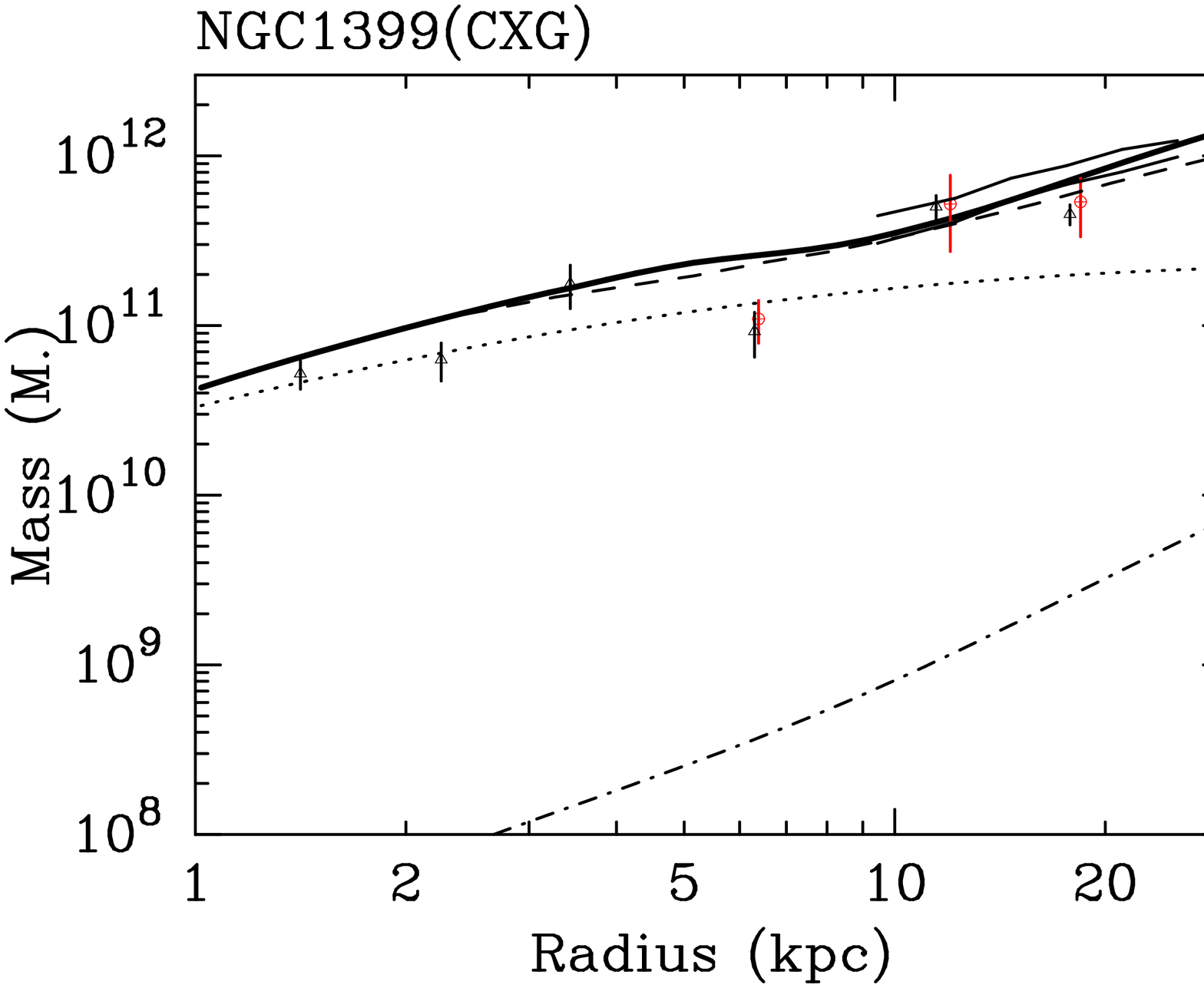}{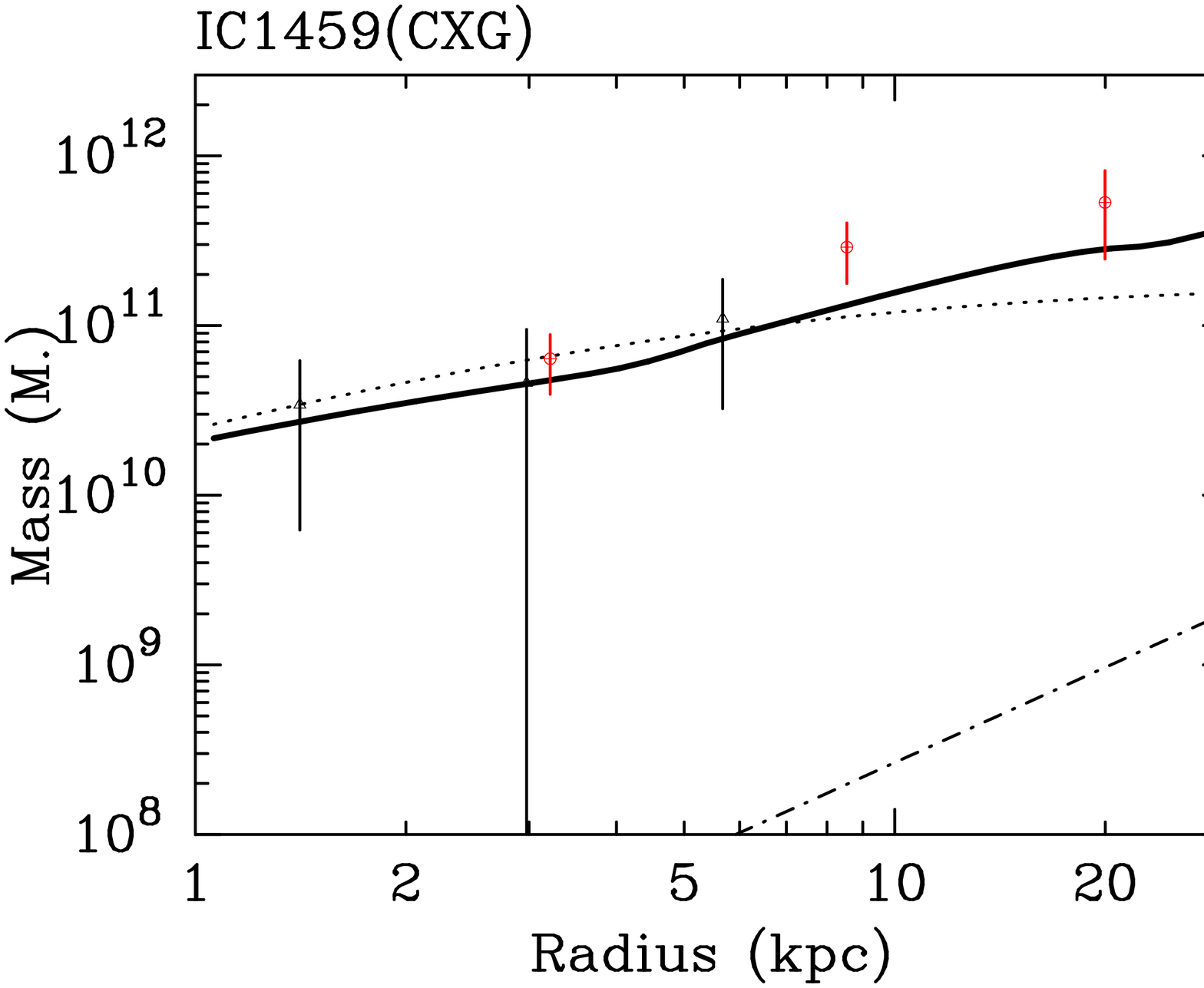}
\plotone{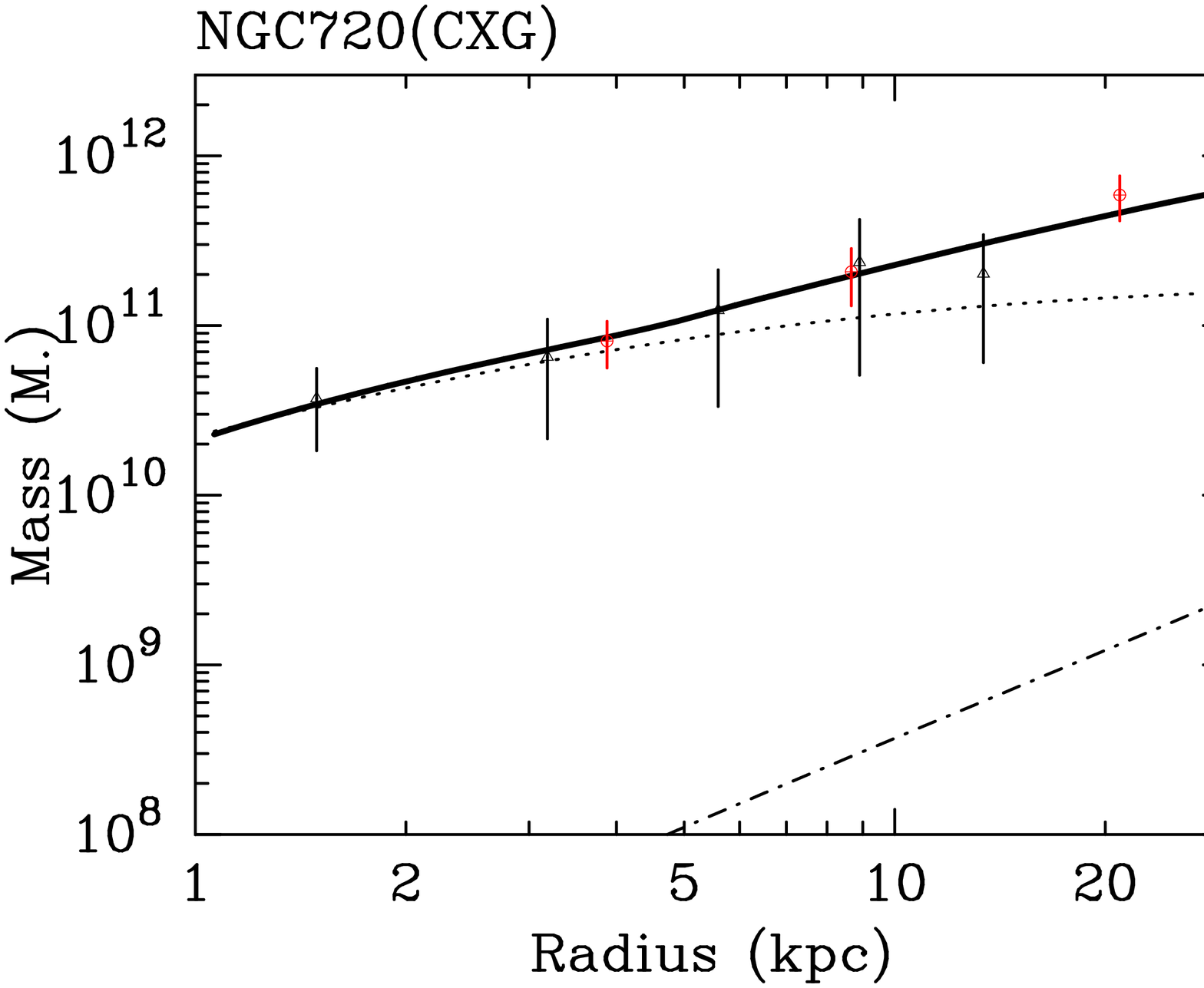}
\vspace*{1cm}
\caption{An example of the mass profiles.
The top-left, top-right, and bottom panels are NGC 1399 (EXG), 
IC 1459 (CXG), and NGC 720 (CXG), respectively.
Triangle and circle points are the total mass profile obtained by analysis 
of {\it Chandra} and {\it XMM-Newton} data, respectively, 
with the deprojection method.
The thick-solid, dot, and dashed-dot line represent the total mass, stellar
 mass, and gas mass, respectively, where the total and gas mass profile
 are obtained by the parameterized analysis and the stellar mass is
 estimated by assuming $M/L_{\rm B}=6(M_{\odot}/L_{\odot})$.
 The dashed line in NGC 1399 is a total mass profile obtained by the
 optical work (Saglia et al. 2000). Two thin-solid lines represent the
 acceptable range of the total mass obtain with ASCA (Ikebe et al. 1996).
\label{n720mass}}
\end{figure}

\begin{figure}
\epsscale{1.0}
\plotone{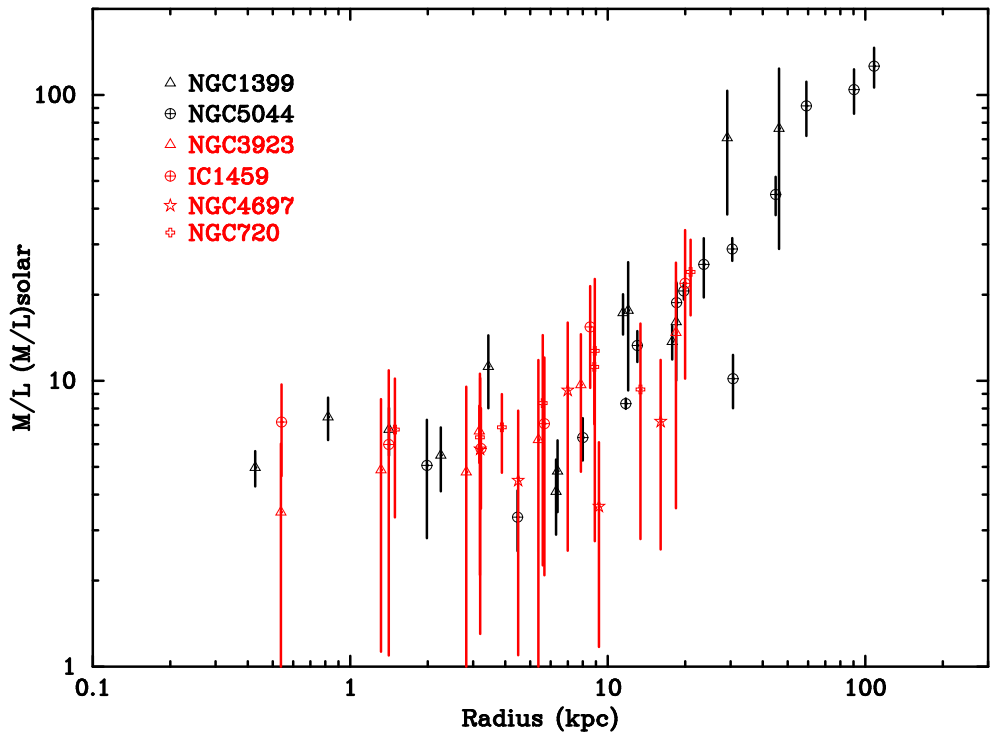}
\vspace*{1cm}
\caption{Radial profile of the mass-to-light ratio for six galaxies, 
obtained by analysis of {\it Chandra} and {\it XMM-Newton} data 
with the deprojection method.
Galaxies with black and red points are EXG and CXG defined in \S3.1.
\label{mlall}}
\end{figure}

In figure \ref{n720mass}, 
we plot an example of the mass profiles of CXGs obtained here.
It can be seen that {\it Chandra} mass profiles at the inner region
smoothly connect to {\it XMM-Newton} profiles at the outer region and
data points of {\it XMM-Newton} give an estimation with a smaller error.
It can be seen that the reults obtained here are almost consistent with those
by the parameterization analyses of the {\it Chandra} data.
The mass of CXGs can be traced beyond 10 kpc, and exceeds much the stellar mass
at the outer region, as well as EXGs.
In figure \ref{mlall}, we plot the radial profiles of the mass-to-light
ratio of six galaxies.
Within 10 kpc, the mass-to-light ratio is
almost constant around 5--8, in good agreement with the results in the
previous subsections.
Beyond 10 kpc, the mass-to-light ratio increases significantly, 
even for CXGs, NGC 3923, NGC 720, and IC 1459.
Although we cannot trace the mass of CXGs
as far away as EXGs, the upturn of the mass-to-light
profiles indicates that the dark matter halo exists around CXGs, 
as well as EXGs.

\begin{figure}
\plottwo{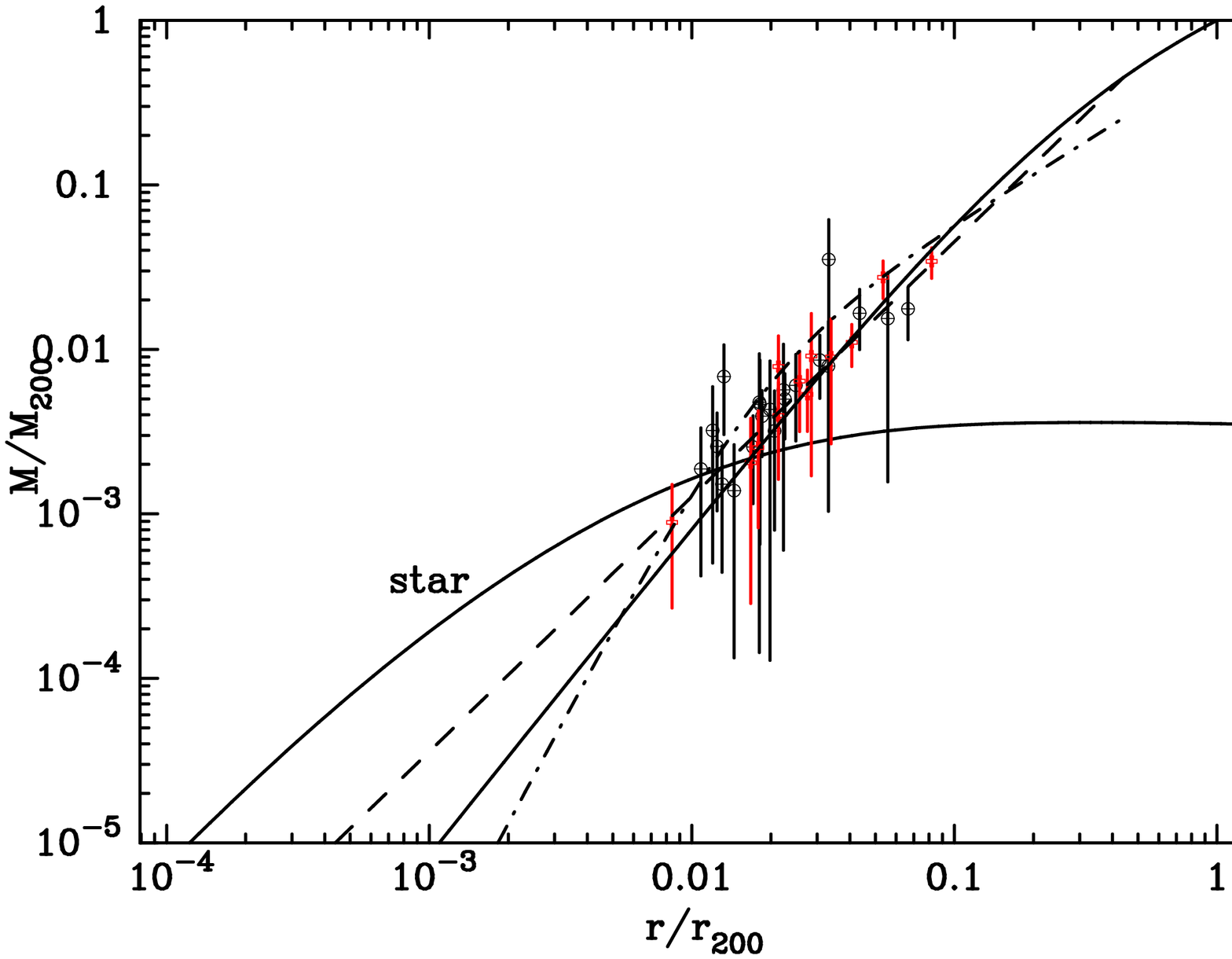}{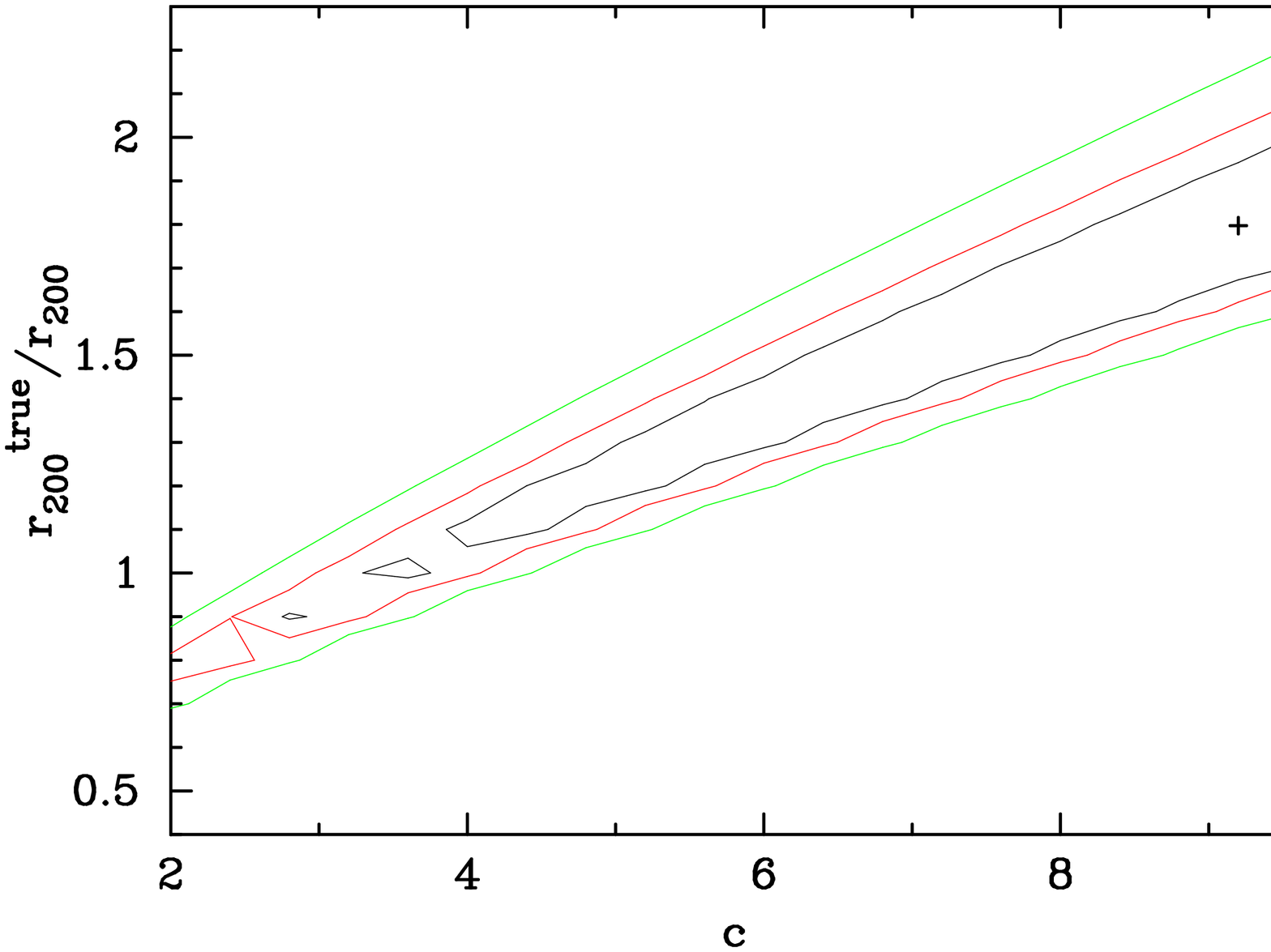}
\caption{Left panel is a scaled mass profile for sample galaxies,
obtained by analysis of {\it Chandra} (open circles) and {\it XMM-Newton} (open triangles) data with the deprojection method.
The data points beyond 10 kpc are plotted anfd fitted, 
The best-fit powerlaw model with the index of 1.66 (dashed line), NFW
 model with $c=3.6$ (solid), and King model with a core radius of
 0.019$r_{200}$ (dot line) are also plotted.
Typical stellar mass profile is also shown, whose parameters are the
same as figure \ref{nfwcmp}.
Right panel is a confidence contour map between the NFW concentration 
parameter $c$ and the $r_{200}^{\rm true}/r_{200}$ for the data points 
in the left panel. Three confidence levels of 68, 90, and 95\% are shown.
\label{nfwfitcmp}}
\end{figure}

Using the data points beyond 10 kpc where the dark matter dominates to
constrain the dark matter profile, as well as the previous subsection,
we obtained $c=3.47\pm0.67$ for NFW profile and the density 
inner slope of $\alpha=-1.34\pm0.33$ for powerlaw profile.
These are consistent with those for only the {\it Chandra} data.
Therefore, we fitted the mass profile, including the {\it Chandra} data
of other galaxies whose {\it XMM-Newton} data are not treated here.
The results are shown in figure \ref{nfwfitcmp} left.
We obtained $c=3.54\pm0.33$ and $\alpha=-1.55\pm0.21$.
Since we assumed the virial radius $r_{200}$ to be as in \S\ref{para-ana},
this value of $c$ indicates the scaling radius
$r_s=r_{200}/c=330\left(kT/{\rm 1 keV}\right)^{0.5}$ kpc.
In most cases, we cannot trace the mass profile up to $r_s$ here, we
cannot determine the convex region in the NFW profile.
This does not allow us to constrain the true virial radius $r_{200}^{\rm
true}$.
At fact, when we fit the mass profile with the NFW model which includes
two parameters, $c$ and $b=r_{200}^{\rm true}/r_{200}$, the best-fit 
values of $c$ varies as shown in figure \ref{nfwfitcmp} right.
We also tried to fit the profile with the King profile \citep{king62},
and the core radius is constrained to be $<0.02r_{200}$.
We cannot rule out the King profile, but the King model cannot fit the
mass profile if the scaling mass profile is the same shape as that of 
galaxy clusters.
Since the stellar mass cannot be completely ignored around 10 kpc, 
the above values of $c$ and $alpha$ are thought to be upper or lower
limit, respectively.
We tried to fit the mass profile with the NFW and powerlaw model, by
considering the typical stellar mass profile which is the same as that
in \S\ref{chandra-dep}.
As a result, we obtained $c=1.83\pm0.53$ for the NFW model and
$\alpha=-0.50\pm0.45$ for the powewlaw model.
Therefore, considering the uncertainties of the assumed stellar mass,
the parameters have a range of $c=2\sim3.5$ or $\alpha=-(0.5\sim1.5)$.

\section{Discussion}

We systematically analyzed {\it Chandra} data of 53 elliptical galaxies, 
and obtained the temperature and ISM density profile to derive the mass 
profile.
Galaxies are mainly classified into two types, based on the temperature and 
ISM density profile.
One type (EXG) has a positive temperature gradient 
toward the outer radius and the ISM density is higher at the outer region.
The other (CXG) has a declining or flat temperature profile and 
the ISM density is lower at the outer region.
These corresponds to X-ray bright elliptical galaxies and X-ray faint ones, 
respectively.
As well as ``X-ray bright/faint'', the separation between EXGs and CXGs
is ambiguous, and they are rather continuously linking with each other.
Both types of galaxies which have been recognized in the previous works
are now clearly distinguished by the temperature and gas density profiles.
Nevertheless, the mass profile is consistent between two types of galaxies.
The mass profile is well scaled by a virial radius $r_{200}$ rather than
 an optical-half radius $r_e$, 
and smoothly connects to those of clusters of galaxies.
This implies that structure formation by dark matter can be described in
 an unified way from rich clusters down to galaxies.

\begin{figure}
\plotone{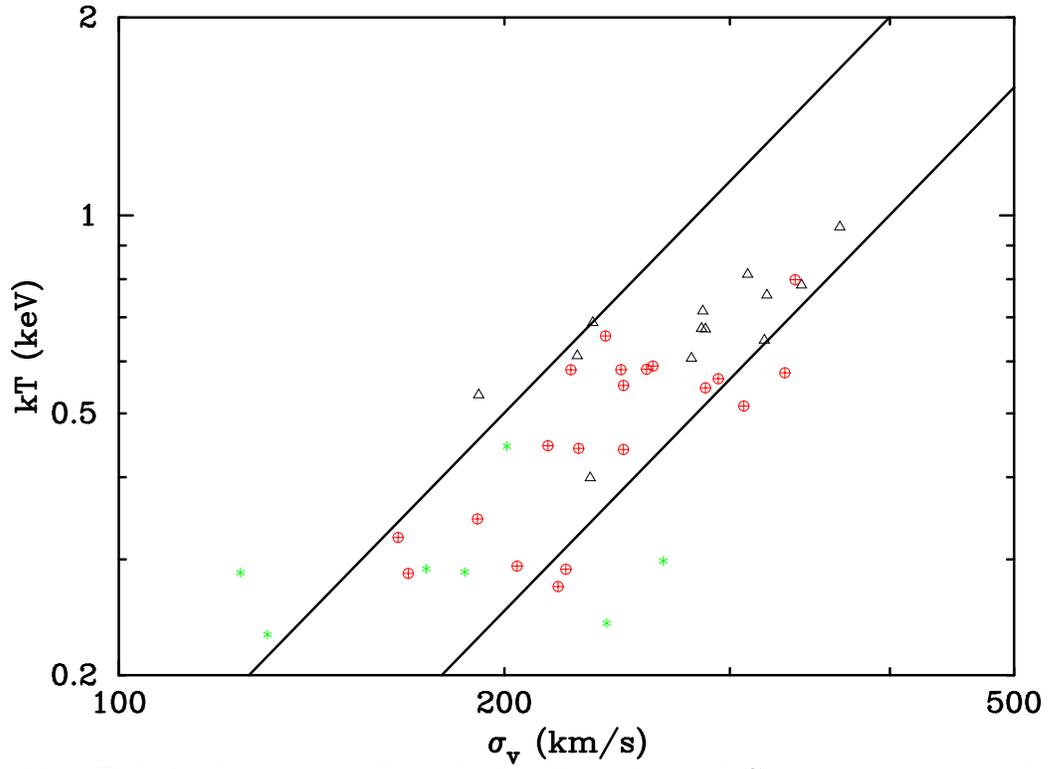}
\caption{Relation between stellar velocity
 dispersion and Gas temperature at the innermost region. The two lines
 correspond to $\beta_{\rm spec}$ of 0.5 and 1.0 for the top and bottom,
 respectively. Symbols are the same as those in figure \ref{r-ktsb}.
\label{vtmll}}
\end{figure}

Higher temperature and higher density of hot gas at the outer region indicates
that EXG is surrounded by the hot intragroup medium.
At fact, the gas mass of some EXGs has found to be comparable to the
stellar mass by the previous works \citep{matsushita98,mulchaey03}.
Lower temperature at the inner region would correspond to the
gravitational potential depth of galaxy itself.
In figure \ref{vtmll}, 
we plot the temperature of the innermost radius against the
stellar velocity dispersion.
Temperature roughly correlates with the velocity dispersion
$\sigma_v$ as $T\propto \sigma_v^{0.5}$ with $\beta_{\rm
spec}=\frac{\mu m_p\sigma_v^2}{kT}=0.5-1.0$,
consistent with the picture that ISM is heated by the stellar motion
\citep{matsushita01}, in other words, gravitational energy.
EXG shows somewhat higher temperature than CXG, indicating that the
temperature of ISM is somewhat affected by the ambient hotter intragroup medium
even at the innermost region.
Therefore, we must pay attention to that ISM properties of EXG do not
necessarily represent galaxy properties.
On the other hand, VCXGs with lower $\sigma_v$
deviate from the typical $T-\sigma_v$ relation of higher $\sigma_v$
CXGs, indicating that their ISM suffers non-gravity heatings 
such as supernovae and stellar mass loss or does not satisfy the
hydrostatic equilibrium.

The mass profile well traces a stellar mass profile with a constant
 mass-to-light ratio of $M/L_{\rm B}=3-10(M_{\odot}/L_{\odot})$ 
at the inner region of (0.001--0.01)$r_{200}$ or (0.1--1)$r_e$.
Therefore, baryon mass dominates in the inner region of elliptical
galaxies; a faction of the star mass is 53--62\% at $0.00885r_{200}$ and
95.5--96.5\% at $0.001r_{200}$ when
we assume the NFW mass profile of $c=3.54\pm0.33$ obtained in
\S\ref{cha-new-dep} and the stellar mass profile in \S\ref{para-ana},
where the optical half radius is $r_e=0.00885r_{200}=8$ kpc, and
$0.001r_{200}$ is equivalent to 0.9 kpc.
However, note that these are not very rigorous limits since composite
mass models are not actually explicitly fit to the data.
$M/L_{\rm B}$ ratio of EXG increases steeply beyond $0.01r_{200}$, 
and thus requires a presence of massive dark matter halo, which
is extended than stars as well as spiral galaxies.
The crossing radius between stars and dark matter is around 10 kpc,
quite similar to that of spiral galaxies with a similar mass scale, and
thus the radial mass profile is not different between ellipticals and
spirals; the difference is presence or absence of stellar disk components.
These indicate that a double structure of gravitational potential in
 X-ray bright elliptical galaxies is mainly due to the different scale
 of mass concentration between stars and dark matter, rather than two
 distinct scales of dark matter.
The mass profile is not well scaled by the optical-half radius $r_e$.
This is concerned with the fact
that $r_e$ scatters widely among galaxies with similar $L_{\rm B}$.
In other words, stellar distribution is different among galaxies with
the same mass, independent of the gravitational potential, as indicated by
\citet{kodama00}.
Connection between individual galaxies and galaxy clusters is also an
interesting issue, and X-ray bright elliptical galaxies are intermediate
systems since they are often located at the center of galaxy groups.
CDM structure formation theories do not give any distinct scales
separating individual galaxies and clusters, while stellar distributions
show a clear separation.
Our results show that dark matter associated with cD galaxy itself is not
distinguished with that of the surround galaxy group.
This implies that cD galaxies evolved together with galaxy groups or
clusters.
{\it XMM-Newton} analyses of 10 galaxy clusters show that the scaled 
mass profile of these objects well trace the NFW profile 
\citep{pointecouteau05}.
Considering it together with our results,
the dark matter profile in elliptical galaxies or galaxy groups 
follow the NFW profile as well as galaxy clusters.

The slope of dark matter profile gives information on the properties of
dark matter, and it is reported to be consistent with the CDM model
for rich clusters \citep{arabadjis02,lewis03,buote04}.
As described in \S\ref{cha-new-dep},
the slope of mass profile at the inner region is $\alpha=-(0.5\sim1.5)$ 
and thus consistent with the NFW profile,
but cannot rule out the self-interacting dark matter model
\citep{spergel00} or adiabatic compression of dark matter \citep{prada04}.
Our constraint is still poor due to the dominant stellar mass at the 
inner region.
On the other hand, mass measurements of low-surface-brightness galaxies
and dwarf irregular galaxies indicate no evidence of DM cusp profile 
at the center region \citep{moore99, cote00, blok01}.
Therefore, measurement of the exact slope of ordinary elliptical
galaxies is important to solve the above problem.
The concentration parameter $c$ for the NFW model is obtained to be 
$c=2\sim3.5$, somewhat smaller than that for 
rich clusters \citep{arabadjis02,lewis03,buote04,pointecouteau05}.
However, if the true virial radius $r_{200}^{\rm true}$ is larger than
the assumed one \citep{evrard96}, a larger value of $c$ is allowed.
Numerical simulations of the CDM halo predict a larger $c$ for
lower-mass systems \citep{navarro97, bullock01}; $c\sim7$ for objects
with the virial mass of $\sim10^{14} M_{\odot}$, corresponding to that
of elliptical galaxies.
Our data allow $c\sim7$ if $r_{200}^{\rm true}/r_{200}\sim1.5$.
Anyway, the mass around elliptical galaxies seems to be less
concentrated than prediction, indicating that the formation epoch is
recent and thus merger evolution of elliptical galaxies is preferred.

For X-ray dim galaxies, we cannot well constrain the dark
 matter content, but find some hints of $M/L_{\rm B}$ increase for CXGs with
 $L_{\rm B}\geq L_{\star}=2.2\times10^{10}L_{\odot}$.
ASCA and ROSAT observations found X-ray emission of such CXGs up to 
(4--8)$r_e$.
Extrapolating the constant temperature profile to this radius, a massive
dark matter becomes necessary for these galaxies with $M/L_{\rm B}\sim50$. 
It is reported that intermediate 
luminous elliptical galaxies with $L_{\rm B}\sim L_{\star}$ do not need 
dark matter within 5$r_e$ \citep{rix97, mendez01, romanowsky03}.
Among galaxies they reported, the mass profile of NGC 4697 is
 analyzed up to $(4-5)r_e$ in this paper.
The $M/L\sim10$ at $5r_e$ is consistent between our results and their
 works.
Additionally, among four CXGs analyzed in subsection \ref{cha-new-dep},
NGC 720 and IC 1459 with $L_{\rm B}\sim L_{\star}$ show a hint of
 the $M/L_{\rm B}$ increase toward the galaxy outskirts.
Therefore, we suggest that ordinary elliptical galaxies commonly contain
a massive dark matter halo.
If CXG has a massive dark matter halo, they could also have an extended
 X-ray emission up to $10r_e$.
Such an indication is also obtained from ASCA observations of galaxies 
and clusters \citep{fukazawa04}.
In our sample, any less massive galaxies with 
$L_{\rm B}<L_{\star}$ do not show enough X-ray emission to trace the dark
 matter, and thus 
studies of less-luminous elliptical galaxies are also needed to
resolve this issue unambiguously.
Since it is probably very faint and cool, X-ray emissivity is too low 
to detect with {\it Chandra}.
Therefore, further improvement of this study will be achieved 
by the low background instrument with a wide field-of-view, 
such as Astro-E2/Suzaku XIS.

\clearpage

\begin{deluxetable}{ccccccccc}
\tabletypesize{\scriptsize}
\tablecaption{List of Sample Elliptical Galaxies \label{tab:sample-chandra}}
\tablewidth{0pt}
\tablehead{
\colhead{Galaxy} & \colhead{SeqID\tablenotemark{a}}   & \colhead{$D$\tablenotemark{b}}   &
\colhead{$L_{\rm B}$\tablenotemark{b}} &
\colhead{$r_e$\tablenotemark{c}}  & \colhead{ACIS} & \colhead{exp. time\tablenotemark{d}} &
\colhead{cts\tablenotemark{d}}\\
\colhead{} & \colhead{} & \colhead{(Mpc)}   & 
\colhead{($log_{10}(L_{\rm B}/L_{\odot}$))}   &
\colhead{(kpc,arcsec)}  & \colhead{} & \colhead{(sec)} &
\colhead{(c/s)}
}
\startdata
IC1262 & 2018 & 142.0 & 10.33 & 8.4 12.22 & ACIS-S & 29797 & 1.345 \\
IC1459 & 2196 & 18.9 & 10.37 & 3.5 38.65 & ACIS-S & 55746 & 0.203 \\
M32 & 2017 &  0.7 &  8.36 & 0.1 38.65 & ACIS-S & 38372 & 0.130 \\
NGC507 & 317 & 67.2 & 10.96 & 25.1 77.11 & ACIS-S & 18364 & 1.507 \\
NGC533 & 2880 & 63.7 & 10.90 & 14.7 47.55 & ACIS-S & 37555 & 0.573 \\
NGC720 & 492 & 20.8 & 10.38 & 4.0 39.55 & ACIS-S & 37691 & 0.438 \\
NGC741 & 2223 & 61.1 & 10.90 & 15.4 52.13 & ACIS-S & 30199 & 0.280 \\
NGC1291 & 795 & 12.1 & 10.33 & 3.2 54.59 & ACIS-S & 37835 & 0.077 \\
NGC1316 & 2022 & 18.1 & 10.93 & 7.1 80.75 & ACIS-S & 23894 & 0.499 \\
NGC1332 & 4372 & 19.7 & 10.27 & 2.7 28.00 & ACIS-S & 56391 & 0.148 \\
NGC1395 & 799 & 20.5 & 10.44 & 4.5 45.41 & ACIS-I & 24423 & 0.251 \\
NGC1399 & 319 & 18.1 & 10.52 & 3.7 42.38 & ACIS-S & 55794 & 2.171 \\
NGC1404 & 2942 & 18.1 & 10.35 & 2.3 26.74 & ACIS-S & 29198 & 1.102 \\
NGC1407 & 791 & 20.6 & 10.58 & 7.2 71.97 & ACIS-S & 46155 & 0.408 \\
NGC1550 & 3186 & 48.5 & 10.33 & 6.0 25.53 & ACIS-I &  9934 & 1.589 \\
NGC1553 & 783 & 14.4 & 10.63 & 4.6 65.63 & ACIS-S & 22725 & 0.290 \\
NGC1600 & 4283 & 60.0 & 11.03 & 13.8 47.55 & ACIS-S & 23575 & 0.235 \\
NGC1700 & 2069 & 56.1 & 10.53 & 3.7 13.71 & ACIS-S & 39293 & 0.149 \\
NGC1705 & 3930 &  4.9 &  8.44 & 0.3 11.67 & ACIS-S & 48208 & 0.034 \\
NGC2434 & 2923 & 14.1 &  9.89 & 2.8 40.47 & ACIS-S & 40819 & 0.214 \\
NGC2681 & 2061 & 10.0 &  9.48 & 4.0 82.46\tablenotemark{\dagger} & ACIS-S & 78928 & 0.031 \\
NGC2865 & 2020 & 36.5 & 10.48 & 2.1 11.67 & ACIS-S & 27871 & 0.160 \\
NGC3115 & 2040 &  8.8 & 10.10 & 1.5 36.07 & ACIS-S & 35646 & 0.068 \\
NGC3377 & 2934 & 10.0 &  9.72 & 1.6 33.66 & ACIS-S & 39593 & 0.003 \\
NGC3379 & 1587 & 10.0 & 10.06 & 1.7 35.25 & ACIS-S & 30680 & 0.062 \\
NGC3585 & 2078 & 16.1 & 10.39 & 3.1 39.55 & ACIS-S & 35245 & 0.068 \\
NGC3607 & 2073 & 19.8 & 10.46 & 6.3 65.63 & ACIS-I & 38448 & 0.098 \\
NGC3923 & 1563 & 17.9 & 10.52 & 4.6 53.35 & ACIS-S & 18887 & 0.262 \\
NGC4125 & 2071 & 25.9 & 10.80 & 7.5 59.86 & ACIS-S & 64086 & 0.196 \\
NGC4261 & 834 & 31.5 & 10.70 & 5.9 38.65 & ACIS-S & 32492 & 0.329 \\
NGC4325 & 3232 & 111.1 & 10.29 & 4.0 7.41\tablenotemark{\dagger} & ACIS-S & 30038 & 1.257 \\
NGC4365 & 2015 & 15.9 & 10.34 & 4.4 57.16 & ACIS-S & 40379 & 0.116 \\
NGC4374 & 803 & 15.9 & 10.57 & 4.2 54.59 & ACIS-S & 28427 & 0.757 \\
NGC4382 & 2016 & 15.9 & 10.64 & 4.2 54.59 & ACIS-S & 39700 & 0.147 \\
NGC4472 & 321 & 15.9 & 10.90 & 8.0 104.02 & ACIS-S & 37071 & 2.064 \\
NGC4494 & 2079 & 21.3 & 10.62 & 4.7 45.41 & ACIS-S & 23602 & 0.152 \\
NGC4552 & 2072 & 15.9 & 10.29 & 2.3 30.00 & ACIS-S & 54370 & 0.366 \\
NGC4555 & 2884 & 90.3 & 10.86 & 1.8 4.01 & ACIS-S & 29718 & 0.129 \\
NGC4621 & 2068 & 15.9 & 10.32 & 3.6 46.46 & ACIS-S & 24787 & 0.060 \\
NGC4636 & 323 & 15.9 & 10.51 & 7.8 101.65 & ACIS-S & 47885 & 2.829 \\
NGC4649 & 785 & 15.9 & 10.73 & 5.7 73.64 & ACIS-S & 32327 & 1.316 \\
NGC4697 & 784 & 15.1 & 10.55 & 5.5 75.36 & ACIS-S & 145366 & 0.108 \\
        & 4727--4729 & \\
NGC5018 & 2070 & 30.2 & 10.57 & 3.7 24.95 & ACIS-S & 30449 & 0.074 \\
NGC5044 & 798 & 30.2 & 10.70 & 11.5 78.91 & ACIS-S & 20415 & 6.250 \\
NGC5102 & 2949 &  4.2 &  9.29 & 0.5 26.13 & ACIS-S & 34162 & 0.092 \\
NGC5171 & 3216 & 99.1 & 10.42 & 12.3 25.53 & ACIS-S & 34624 & 0.258 \\
NGC5253 & 2032 &  3.6 &  8.96 & 0.4 22.76 & ACIS-S & 55791 & 0.152 \\
NGC5846 & 788 & 22.9 & 10.66 & 7.0 62.68 & ACIS-S & 24680 & 1.526 \\
NGC5866 & 2879 & 13.2 & 10.32 & 2.6 40.47 & ACIS-S & 33685 & 0.033 \\
NGC6861 & 3190 & 34.7 & 10.42 & 3.8 22.76 & ACIS-I & 21384 & 0.295 \\
NGC6868 & 3191 & 34.7 & 10.58 & 6.5 38.65 & ACIS-I & 23414 & 0.272 \\
NGC7618 & 802 & 74.8 & 10.06 & 4.0 11.00\tablenotemark{\dagger} & ACIS-S & 14804 & 0.763 \\
NGC7619 & 2074 & 40.0 & 10.58 & 6.2 32.15 & ACIS-I & 26592 & 0.307 \\
\enddata
\tablenotetext{a}{Sequence number of the {\it Chandra} data}
\tablenotetext{b}{$D$ is a distance to the galaxy. $L_{\rm B}$ is an
 optical blue luminosity. These are taken from \citet{osullivan01} and
 NED (NASA/IPAC EXTRAGALACTIC DATABASE).}
\tablenotetext{c}{$r_e$ is an optical half radius (effective radius)
 taken from \citet{faber89}. Values indicated by $\dagger$ are assumed
 to be $r_e=4$ kpc, due to no available data.}
\tablenotetext{d}{Exposure and source count rate in the ACIS-S 3 or ACIS-I 0-3.}
\end{deluxetable}

\begin{deluxetable}{ccc}
\tabletypesize{\scriptsize}
\tablecaption{List of Sample Elliptical Galaxies observed with {\it XMM-Newton} \label{tab:sample-newton}}
\tablewidth{0pt}
\tablehead{
\colhead{Galaxy} & \colhead{SeqID\tablenotemark{a}}  & \colhead{exp. time\tablenotemark{b}} \\
\colhead{} & \colhead{} & \colhead{(ksec)} 
}
\startdata
IC 1459 & 0135980201 & 30,30,28 \\\
NGC 720 & 0112300101 & 20,20,18 \\
NGC 1399 & 0012830101 & 23,23,-- \\
NGC 3923 & 0027340101 & 33,33,31 \\
NGC 4697 & 0153450101 & 46,47,-- \\
NGC 5044 & 0037950101 & 22,22,17 \\
\enddata
\tablenotetext{a}{Sequence number of the {\it XMM-Newton} data}
\tablenotetext{b}{Exposure time of the EPIC-MOS1, -MOS2, and -PN.}
\end{deluxetable}

\begin{deluxetable}{cccccccc}
\tabletypesize{\scriptsize}
\tablecaption{Results of overall spectral analyses of Sample Elliptical Galaxies \label{table:spec}}
\tablewidth{0pt}
\tablehead{
\colhead{Galaxy} & \colhead{Radius\tablenotemark{a}}   
& \colhead{Galactic $N_{\rm H}$\tablenotemark{b}}  & \colhead{$kT$\tablenotemark{b}}  & \colhead{$L_{\rm X, soft}$\tablenotemark{b}}
& \colhead{$L_{\rm X, hard}$\tablenotemark{b}}
& \colhead{$\chi^2$/d.o.f.\tablenotemark{b}} &
 \colhead{type\tablenotemark{d}} \\
\colhead{} & \colhead{(arcsec)} & \colhead{($10^{20}$ cm$^{-2}$)} &
\colhead{(keV)}   & \colhead{(erg s$^{-1}$)} & \colhead{(erg s$^{-1}$)} & \colhead{} & \colhead{} 
}
\startdata
IC1262 & 0--240 & 2.43\tablenotemark{c} & $1.52\pm0.03$ & $9.2\times10^{42}$ & $5.3\times10^{42}$ & 1.56 & EXG \\ 
IC1459 & 0--240 & 1.17 & $0.47\pm0.02$ & $1.2\times10^{40}$ & $4.1\times10^{40}$ & 1.58 & CXG \\ 
M32 & 12--100 & 6.38 & $0.34\pm0.07$ & $1.1\times10^{36}$ & $6.6\times10^{36}$ & 0.91 & VCXG \\ 
NGC507 & 0--240 & 5.23\tablenotemark{c} & $1.08\pm0.01$ & $2.0\times10^{42}$ & $1.6\times10^{42}$ & 2.18 & EXG \\ 
NGC533 & 0--240 & 3.07\tablenotemark{c} & $0.96\pm0.01$ & $8.2\times10^{41}$ & $2.8\times10^{41}$ & 2.06 & EXG \\ 
NGC720 & 0--240 & 1.58 & $0.54\pm0.02$ & $3.7\times10^{40}$ & $6.2\times10^{40}$ & 2.17 & CXG \\ 
NGC741 & 0--240 & 4.44 & $0.96\pm0.02$ & $2.7\times10^{41}$ & $2.5\times10^{41}$ & 1.60 & EXG \\ 
NGC1291 & 0--90 & 2.24 & $0.31\pm0.02$ & $1.7\times10^{39}$ & $4.7\times10^{39}$ & 1.66 & VCXG \\ 
NGC1316 & 0--240 & 2.13 & $0.60\pm0.01$ & $3.9\times10^{40}$ & $3.3\times10^{40}$ & 2.45 & CXG \\ 
NGC1332 & 0--240 & 2.30 & $0.54\pm0.02$ & $1.9\times10^{40}$ & $1.9\times10^{40}$ & 2.06 & CXG \\ 
NGC1395 & 0--240 & 1.94 & $0.74\pm0.03$ & $2.6\times10^{40}$ & $3.0\times10^{40}$ & 1.89 & CXG \\ 
NGC1399 & 0--240 & 1.49\tablenotemark{c} & $1.17\pm0.01$ & $2.7\times10^{41}$ & $2.5\times10^{39}$ & 6.02 & EXG \\ 
NGC1404 & 0--240 & 1.51\tablenotemark{c} & $0.52\pm0.01$ & $1.2\times10^{41}$ & $3.7\times10^{40}$ & 1.80 & CXG \\ 
NGC1407 & 0--240 & 5.42\tablenotemark{c} & $0.77\pm0.01$ & $5.1\times10^{40}$ & $4.0\times10^{40}$ & 1.70 & EXG \\ 
NGC1550 & 0--240 & 11.24 & $1.30\pm0.01$ & $1.9\times10^{42}$ & $2.4\times10^{41}$ & 1.47 & EXG \\ 
NGC1553 & 6--120 & 1.50 & $0.39\pm0.01$ & $6.6\times10^{39}$ & $7.5\times10^{39}$ & 1.97 & CXG \\ 
NGC1600 & 0--240 & 4.86 & $1.02\pm0.03$ & $1.9\times10^{41}$ & $3.4\times10^{41}$ & 1.33 & EXG \\ 
NGC1700 & 0--120 & 4.76 & $0.38\pm0.02$ & $7.2\times10^{40}$ & $7.6\times10^{40}$ & 1.96 & EXG \\ 
NGC1705 & 0--60 & 3.85 & $0.19\pm0.02$ & $4.8\times10^{37}$ & $7.9\times10^{37}$ & 1.02 & VCXG \\ 
NGC2434 & 0--120 & 12.23 & $0.38\pm0.03$ & $3.9\times10^{39}$ & $8.7\times10^{39}$ & 1.37 & CXG \\ 
NGC2681 & 0--34 & 2.48 & $0.36\pm0.02$ & $5.9\times10^{38}$ & $5.5\times10^{38}$ & 1.47 & VCXG \\ 
NGC2865 & 0--80 & 6.31 & $0.33\pm0.10$ & $2.0\times10^{39}$ & $1.3\times10^{40}$ & 1.27 & CXG \\ 
NGC3115 & 0--240 & 4.61 & $0.32\pm0.02$ & $10.0\times10^{38}$ & $2.3\times10^{39}$ & 0.84 & VCXG \\ 
NGC3377 & 0--30 & 2.78 & $0.25\pm0.05$ & $9.3\times10^{37}$ & $5.1\times10^{38}$ & 1.16 & VCXG \\ 
NGC3379 & 12--18 & 2.78 & $0.52\pm0.30$ & $4.1\times10^{37}$ & $1.2\times10^{38}$ & 0.85 & VCXG \\ 
NGC3585 & 0--150 & 5.60 & $0.31\pm0.02$ & $2.8\times10^{39}$ & $2.3\times10^{39}$ & 1.19 & CXG \\ 
NGC3607 & 0--240 & 1.48 & $0.58\pm0.02$ & $1.5\times10^{40}$ & $1.2\times10^{39}$ & 0.98 & CXG \\ 
NGC3923 & 0--240 & 6.30 & $0.42\pm0.02$ & $2.3\times10^{40}$ & $2.2\times10^{40}$ & 1.29 & CXG \\ 
NGC4125 & 0--240 & 1.82 & $0.39\pm0.01$ & $4.5\times10^{40}$ & $1.4\times10^{40}$ & 1.38 & CXG \\ 
NGC4261 & 0--240 & 1.58 & $0.68\pm0.01$ & $7.4\times10^{40}$ & $9.8\times10^{40}$ & 1.54 & CXG \\ 
NGC4325 & 0--240 & 2.14\tablenotemark{c} & $0.83\pm0.01$ & $7.0\times10^{42}$ & $1.1\times10^{40}$ & 1.75 & EXG \\ 
NGC4365 & 0--240 & 1.61 & $0.64\pm0.03$ & $6.2\times10^{39}$ & $9.2\times10^{39}$ & 1.33 & CXG \\ 
NGC4374 & 0--240 & 2.78 & $0.66\pm0.01$ & $4.5\times10^{40}$ & $3.6\times10^{40}$ & 1.36 & CXG \\ 
NGC4382 & 0--240 & 2.50 & $0.37\pm0.01$ & $1.2\times10^{40}$ & $5.5\times10^{39}$ & 1.31 & CXG \\ 
NGC4472 & 0--240 & 1.62\tablenotemark{c} & $0.85\pm0.01$ & $1.8\times10^{41}$ & $4.8\times10^{40}$ & 5.02 & EXG \\ 
NGC4494 & 0--40 & 1.50 & $0.33\pm0.07$ & $9.7\times10^{38}$ & $9.0\times10^{39}$ & 1.28 & VCXG \\ 
NGC4552 & 0--120 & 2.56 & $0.52\pm0.01$ & $2.3\times10^{40}$ & $9.6\times10^{39}$ & 1.36 & CXG \\ 
NGC4555 & 0--240 & 1.33 & $0.87\pm0.04$ & $3.5\times10^{41}$ & $2.4\times10^{41}$ & 1.28 & EXG \\ 
NGC4621 & 0--80 & 2.17 & $0.25\pm0.05$ & $5.3\times10^{38}$ & $3.8\times10^{39}$ & 0.60 & VCXG \\ 
NGC4636 & 0--240 & 1.82\tablenotemark{c} & $0.67\pm0.01$ & $2.3\times10^{41}$ & $5.0\times10^{39}$ & 5.51 & EXG \\ 
NGC4649 & 0--240 & 2.13\tablenotemark{c} & $0.75\pm0.01$ & $1.1\times10^{41}$ & $2.8\times10^{40}$ & 4.57 & EXG \\ 
NGC4697 & 0--240 & 2.14 & $0.34\pm0.01$ & $7.0\times10^{39}$ & $1.1\times10^{39}$ & 1.62 & CXG \\ 
NGC5018 & 0--150 & 7.14 & $0.32\pm0.02$ & $1.1\times10^{40}$ & $1.1\times10^{40}$ & 1.21 & CXG \\ 
NGC5044 & 0--240 & 4.94\tablenotemark{c} & $0.83\pm0.01$ & $1.9\times10^{42}$ & $9.9\times10^{40}$ & 3.79 & EXG \\ 
NGC5102 & 0--240 & 4.32 & $0.23\pm0.01$ & $5.2\times10^{38}$ & $1.1\times10^{38}$ & 1.26 & VCXG \\ 
NGC5171 & 0--40 & 1.94 & $0.63\pm0.06$ & $3.6\times10^{40}$ & $4.3\times10^{40}$ & 1.59 & CXG \\ 
NGC5253 & 0--120 & 3.88 & $0.30\pm0.01$ & $2.0\times10^{38}$ & $3.2\times10^{38}$ & 1.57 & VCXG \\ 
NGC5846 & 0--240 & 4.24\tablenotemark{c} & $0.66\pm0.01$ & $2.6\times10^{41}$ & $3.1\times10^{40}$ & 2.93 & EXG \\ 
NGC5866 & 0--74 & 1.47 & $0.33\pm0.02$ & $1.5\times10^{39}$ & $1.4\times10^{39}$ & 1.60 & VCXG \\ 
NGC6861 & 0--240 & 5.01 & $0.94\pm0.03$ & $9.3\times10^{40}$ & $5.8\times10^{40}$ & 1.39 & EXG \\ 
NGC6868 & 0--240 & 4.96 & $0.70\pm0.02$ & $1.2\times10^{41}$ & $1.6\times10^{40}$ & 1.14 & EXG \\ 
NGC7618 & 0--240 & 11.93 & $0.78\pm0.01$ & $8.9\times10^{41}$ & $1.5\times10^{42}$ & 2.66 & EXG \\ 
NGC7619 & 0--240 & 5.04 & $0.94\pm0.02$ & $1.9\times10^{41}$ & $3.4\times10^{40}$ & 1.19 & CXG \\ 
\enddata
\tablenotetext{a}{The inner and outer radii to integrate the spectrum.}
\tablenotetext{b}{The fitting results with APEC model. $kT$ is a
 temperature, $L_{\rm X, soft}$ is an X-ray luminosity of the thermal
 emission in the range of 0.2--5 keV, and $L_{\rm X, hard}$ is an X-ray 
 luminosity of the hard emission in the range of 2--10 keV. }
\tablenotetext{c}{The column density $N_{\rm H}$ is left to be a free parameter.}
\tablenotetext{d}{EXG or CXG is a galaxy with the hot gas electron
 density at 10 kpc to be more or less than $2\times10^{-3}$ cm$^{-3}$,
 respectively. VCXG is a galaxy whose X-ray emission is limited within 10
 kpc.}
\end{deluxetable}

\begin{deluxetable}{cccccccccc}
\tabletypesize{\scriptsize}
\tablecaption{Results of analyses of Sample Elliptical Galaxies \label{table:results}}
\tablewidth{0pt}
\tablehead{
\colhead{Galaxy} & \colhead{$kT_i$\tablenotemark{a}}   
& \colhead{$kT_o$\tablenotemark{a}}   & \colhead{$n_{\rm t-r}$\tablenotemark{b}}
& \colhead{$n_{\beta}$\tablenotemark{b}} &
 \colhead{$n_{\rm depro}$\tablenotemark{c}} & \colhead{$R_{\rm max}$\tablenotemark{d}} & 
\colhead{$M/L_{\rm B}$\tablenotemark{e}} & \colhead{$n_{\rm e,10kpc}$\tablenotemark{f}} &
\colhead{type\tablenotemark{g}} \\
\colhead{} & \colhead{(keV, arcsec)} & 
\colhead{(keV)}   & \colhead{} & \colhead{} & \colhead{} & \colhead{(kpc)} &
\colhead{$(M_{\odot}/L_{\odot})$}  & \colhead{($10^{-3}$cm$^{-3}$)} & \colhead{}
}
\startdata
IC1262 & 1.04$\pm$0.26 (9) & 1.21$\pm$0.27 & 2\tablenotemark{\dagger} & 2 & 9 & $>$273.94 & 31.61 & 12.40 & EXG \\
IC1459 & 0.61$\pm$0.11 (5) & 0.38$\pm$0.22 & 2\tablenotemark{\dagger} & 1 & 7 & 36.41 & 4.33 & 1.42 & CXG \\
M32 & 0.34$\pm$0.07 (56) & --- & 0 & 1 & --- & 0.35 & 20.61 & --- & VCXG \\
NGC507 & 0.99$\pm$0.29 (10) & 1.07$\pm$0.08 & 2 & 2 & 6 & $>$129.58 & 17.79 & 7.44 & EXG \\
NGC533 & 0.59$\pm$0.07 (7) & 1.28$\pm$0.22 & 2\tablenotemark{\dagger} & 3 & 9 & $>$122.81 & 22.71 & 9.39 & EXG \\
NGC720 & 0.64$\pm$0.18 (7) & 0.49$\pm$0.28 & 2\tablenotemark{\dagger} & 2 & 12 & $>$40.11 & 7.30 & 1.80 & CXG \\
NGC741 & 0.61$\pm$0.14 (7) & 1.32$\pm$0.19 & 2 & 2 & 6 & $>$117.82 & 20.80 & 4.62 & EXG \\
NGC1291 & 0.31$\pm$0.02 (45) & --- & 0 & 1 & --- & 5.22 & 5.12 & --- & VCXG \\
NGC1316 & 0.61$\pm$0.09 (7) & 0.44$\pm$0.21 & 2\tablenotemark{\dagger} & 1 & 9 & 17.55 & 3.47 & 1.62 & CXG \\
NGC1332 & 0.59$\pm$0.07 (7) & 0.41$\pm$0.28 & 2 & 2 & 5 & 27.27 & 7.67 & 0.85 & CXG \\
NGC1395 & 0.59$\pm$0.20 (24) & 0.68$\pm$0.19 & 2 & 1 & 4 & 69.77 & 8.38 & 0.28 & CXG \\
NGC1399 & 0.84$\pm$0.03 (5) & 1.19$\pm$0.05 & 2\tablenotemark{\dagger} & 2 & 16 & $>$34.93 & 10.95 & 3.96 & EXG \\
NGC1404 & 0.56$\pm$0.05 (7) & 0.45$\pm$0.08 & 2\tablenotemark{\dagger} & 1 & 4 & $>$34.93 & 5.51 & 0.52 & CXG \\
NGC1407 & 0.76$\pm$0.11 (7) & 0.95$\pm$0.13 & 2\tablenotemark{\dagger} & 3 & 8 & $>$39.75 & 21.21 & 2.01 & EXG \\
NGC1550 & 1.15$\pm$0.30 (18) & 1.19$\pm$0.12 & 2 & 3 & 9 & $>$172.47 & 23.95 & 18.26 & EXG \\
NGC1553 & 0.27$\pm$0.19 (19) & 0.37$\pm$0.12 & 2 & 1 & 5 & $>$27.87 & 3.74 & 1.22 & CXG \\
NGC1600 & 0.83$\pm$0.10 (12) & 1.56$\pm$0.38 & 2 & 2 & 4 & $>$115.68 & 10.58 & 3.38 & EXG \\
NGC1700 & 0.44$\pm$0.19 (5) & 0.33$\pm$0.12 & 2 & 2 & 2 & 32.36 & 6.26 & 2.04 & EXG \\
NGC1705 & 0.19$\pm$0.02 (30) & --- & 0 & 1 & --- & 1.39 & 5.31 & --- & VCXG \\
NGC2434 & 0.32$\pm$0.19 (34) & 0.29$\pm$0.28 & 0 & 2 & 5 & $>$27.12 & 5.32 & 0.90 & CXG \\
NGC2681 & 0.36$\pm$0.02 (17) & --- & 0 & 2 & --- & 1.60 & --- & --- & VCXG \\
NGC2865 & 0.33$\pm$0.10 (40) & --- & 0 & 1 & --- & 62.79 & 2.41 & 0.24 & CXG \\
NGC3115 & 0.32$\pm$0.02 (120) & --- & 0 & 1 & --- & 4.24 & 4.04 & --- & VCXG \\
NGC3377 & 0.25$\pm$0.05 (15) & --- & 0 & 1 & --- & 1.41 & --- & --- & VCXG \\
NGC3379 & 0.52$\pm$0.30 (15) & --- & 0 & 1 & --- & 0.82 & 13.93 & --- & VCXG \\
NGC3585 & 0.31$\pm$0.02 (75) & --- & 0 & 2 & --- & 11.60 & 2.75 & 0.76 & CXG \\
NGC3607 & 0.63$\pm$0.28 (29) & 0.53$\pm$0.31 & 2 & 2 & --- & 28.55 & 4.50 & 1.37 & CXG \\
NGC3923 & 0.49$\pm$0.16 (7) & 0.40$\pm$0.15 & 2 & 2 & 4 & $>$34.44 & 6.85 & 1.26 & CXG \\
NGC4125 & 0.44$\pm$0.24 (7) & 0.36$\pm$0.06 & 2 & 1 & 9 & 37.46 & 4.82 & 1.89 & CXG \\
NGC4261 & 0.56$\pm$0.08 (5) & 1.10$\pm$0.58 & 2\tablenotemark{\dagger} & 2 & 5 & $>$60.71 & 9.52 & 1.45 & CXG \\
NGC4325 & 0.65$\pm$0.10 (7) & 0.92$\pm$0.04 & 2 & 3 & 9 & $>$214.20 & 6.55 & 25.02 & EXG \\
NGC4365 & 0.34$\pm$0.23 (10) & 0.48$\pm$0.33 & 2 & 3 & --- & 30.70 & 8.10 & 0.82 & CXG \\
NGC4374 & 0.57$\pm$0.09 (5) & 0.72$\pm$0.12 & 2\tablenotemark{\dagger} & 1 & 11 & 15.43 & 7.27 & 0.63 & CXG \\
NGC4382 & 0.36$\pm$0.25 (12) & 0.36$\pm$0.36 & 2 & 2 & 3 & 22.99 & 2.72 & 1.45 & CXG \\
NGC4472 & 0.65$\pm$0.04 (7) & 0.94$\pm$0.04 & 2\tablenotemark{\dagger} & 2 & 10 & $>$30.70 & 6.93 & 5.65 & EXG \\
NGC4494 & 0.26$\pm$0.14 (20) & 0.29$\pm$0.11 & 0 & 1 & --- & 5.05 & 4.64 & --- & VCXG \\
NGC4552 & 0.57$\pm$0.07 (7) & 0.42$\pm$0.16 & 2\tablenotemark{\dagger} & 2 & 9 & 143.89 & 7.56 & 0.95 & CXG \\
NGC4555 & 0.68$\pm$0.16 (5) & 1.12$\pm$0.48 & 2 & 2 & --- & 119.93 & 1.20 & 3.57 & EXG \\
NGC4621 & 0.25$\pm$0.05 (40) & --- & 0 & 1 & --- & 4.55 & 4.80 & --- & VCXG \\
NGC4636 & 0.53$\pm$0.06 (7) & 0.65$\pm$0.01 & 2 & 2 & 30 & $>$30.70 & 15.74 & 5.70 & EXG \\
NGC4649 & 0.86$\pm$0.04 (5) & 0.80$\pm$0.07 & 2\tablenotemark{\dagger} & 1 & 5 & $>$30.70 & 9.15 & 2.57 & EXG \\
NGC4697 & 0.31$\pm$0.07 (23) & 0.32$\pm$0.06 & 0 & 1 & 6 & $>$29.20 & 3.56 & 0.88 & CXG \\
NGC5018 & 0.32$\pm$0.02 (75) & --- & 0 & 1 & --- & 21.80 & 3.21 & 0.99 & CXG \\
NGC5044 & 0.68$\pm$0.08 (7) & 0.81$\pm$0.02 & 2 & 2 & 17 & $>$58.24 & 15.53 & 20.47 & EXG \\
NGC5102 & 0.23$\pm$0.01 (120) & --- & 0 & 1 & --- & 0.58 & 2.65 & --- & VCXG \\
NGC5171 & 0.77$\pm$0.48 (48) & 0.65$\pm$0.14 & 0 & 2 & 4 & 18.73 & 18.19 & 1.71 & CXG \\
NGC5253 & 0.33$\pm$0.14 (9) & --- & 2 & 1 & 5 & 0.69 & 11.59 & --- & VCXG \\
NGC5846 & 0.71$\pm$0.14 (7) & 0.64$\pm$0.05 & 2\tablenotemark{\dagger} & 2 & 9 & $>$44.18 & 6.70 & 7.17 & EXG \\
NGC5866 & 0.33$\pm$0.02 (37) & --- & 0 & 1 & --- & 4.66 & 4.46 & --- & VCXG \\
NGC6861 & 0.68$\pm$0.10 (18) & 0.98$\pm$0.15 & 2 & 2 & 4 & $>$123.98 & 13.32 & 2.08 & EXG \\
NGC6868 & 0.68$\pm$0.25 (35) & 0.66$\pm$0.14 & 2 & 2 & 4 & $>$119.95 & 5.71 & 2.35 & EXG \\
NGC7618 & 0.80$\pm$0.19 (12) & 0.70$\pm$0.07 & 2 & 3 & 6 & $>$144.22 & 26.45 & 9.66 & EXG \\
NGC7619 & 0.78$\pm$0.14 (8) & 0.88$\pm$0.11 & 2 & 2 & 6 & 144.56 & 13.77 & 1.04 & CXG \\
\enddata
\tablenotetext{a}{$T_i$ and $T_o$ is the gas temperature at the
 innermost radial bin and 10 kpc in the radial temperature profile. When
 the number of the radial bin is smaller than 5, we adopt the
 temperature in table \ref{table:spec} as $T_i$. ``---'' for $T_o$
 indicates that $R_{\rm max}$ is smaller than 10 kpc, or the number of
 the radial bin is smaller than 5. The parentheses in the $T_i$ column
 represent the averaged annular radius (in unit of arcsec) 
where $T_i$ is obtained.}
\tablenotetext{b}{$n_{\rm t-r}$ is an order of the polynomial function
 modeling the temperature profile. $\dagger$ indicates that the profile
 is jointed between the inner $50''$ and outer $50''-$ region. See text
 in detail. $n_{\beta}$ is a number of the
 $\beta$ model component which fits the radial X-ray surface brightness 
 profile. The modeling of the AGN point source is not included in this number.}
\tablenotetext{c}{A number of deprojected spectra on {\it Chandra} data.}
\tablenotetext{d}{The maxiums detection radius (kpc).}
\tablenotetext{e}{The mass-to-light ratio $M/L_{\rm B}$ at an effective
 radius $r_e$, obtained by the parameterization analysis in \S\ref{para-ana}.}
\tablenotetext{f}{The hot gas electron density at 10 kpc.}
\tablenotetext{g}{EXG or CXG is a galaxy with the hot gas electron
 density at 10 kpc to be more or less than $2\times10^{-3}$ cm$^{-3}$,
 respectively. VCXG is a galaxy whose X-ray emission is limited within 10
 kpc.}
\end{deluxetable}

\begin{deluxetable}{cccccc}
\tabletypesize{\scriptsize}
\tablecaption{Comparison of the total mass between the optical results and ours \label{table:nfwcmp}}
\tablewidth{0pt}
\tablehead{
\colhead{} & \colhead{Optical} & \colhead{X-ray} & \colhead{X-ray} & \colhead{} & \colhead{}  \\
\colhead{Galaxy} 
& \colhead{$M_{\rm tot}$\tablenotemark{a}}   
& \colhead{$M_{\rm tot}$\tablenotemark{b}}
& \colhead{$(M/L_{\rm B})$\tablenotemark{b}} 
& \colhead{Radius\tablenotemark{c}} & \colhead{Reference\tablenotemark{d}} \\
\colhead{} & 
\colhead{($M_{\odot}$)} &
\colhead{($M_{\odot}$)} & 
\colhead{$(M_{\odot}/L_{\odot})$} & 
\colhead{(arcsec, $r_e$)} & \colhead{} 
}
\startdata
NGC 1316 & $3.2\times10^{11}$ & $3.9\times10^{11}$ & 5.6 & 200, 2.5 & 2 \\
NGC 1399 & $9.9\times10^{10}$ & $1.9\times10^{11}$ & 10.9 & 42, 1.0 & 1 \\
         & $2.1\times10^{11}$ & $2.5\times10^{11}$ & 11.6 & 60, 1.4 & 3 \\
NGC 2434 & $1.6\times10^{10}$ & $1.3\times10^{10}$ & 4.9 & 24, 0.6 & 1 \\
         & $1.6\times10^{10}$ & $1.3\times10^{10}$ & 4.9 & 24, 0.6 & 4 \\
NGC 3379 & $1.2\times10^{10}$ & $8.2\times10^{10}$ & 13.9 & 35, 1.0 & 1 \\
NGC 4374 & $1.2\times10^{11}$ & $1.4\times10^{11}$ & 7.3 & 57, 1.0 & 1 \\
NGC 4697 & $1.3\times10^{11}$ & $8.7\times10^{10}$ & 3.8 & 95, 1.2 & 5 \\
NGC 5846 & $1.3\times10^{11}$ & $2.3\times10^{11}$ & 8.4 & 83, 1.3 & 1 \\
\enddata
\tablenotetext{a}{The total mass obtained by the optical works. Typical
 measurement error is 10--20\%, but systematic errors in converting the
 mass-to-light ratio in Kronawitter et al. (2000) into the mass are 
 possibly $\sim$50\%.}
\tablenotetext{b}{The total mass and mass-to-light ratio in this
 work. Typical error is 30\%.}
\tablenotetext{c}{The radius at which the mass-to-light ratio or the
 total mass is calculated.}
\tablenotetext{d}{1: Kronawitter et al. (2000), 2: Arnaboldi et
 al. (1998), 3: Saglia et al. (2000), 4: Rix et al. (1997), 5: M\'endez et al. (2001)}
\end{deluxetable}

\end{document}